\documentclass[12pt]{article}

\usepackage{latexsym}
\usepackage{amsmath}
\usepackage{verbatim}
\usepackage{amssymb}
\usepackage{multirow}
\usepackage[T1]{fontenc}
\usepackage{charter}

\usepackage{physics}

\usepackage{fancyhdr}
\usepackage{hyperref}

\usepackage{lastpage}

\usepackage{xcolor}
\usepackage{tikz}
\usepackage{hvlogos}
\usepackage[
top    = 2.75cm,
bottom = 2.75cm, left=20mm,right=20mm]{geometry}

\numberwithin{equation}{subsection}

\catcode`\@=11
\def\marginnote#1{}
\newcount\hour
\newcount\minute
\newtoks\amorpm
\hour=\time\divide\hour by60
\minute=\time{\multiply\hour by60 \global\advance\minute by-\hour}
\edef\standardtime{{\ifnum\hour<12 \global\amorpm={am}%
        \else\global\amorpm={pm}\advance\hour by-12 \fi
        \ifnum\hour=0 \hour=12 \fi
        \number\hour:\ifnum\minute<10 0\fi\number\minute\the\amorpm}}
\edef\militarytime{\number\hour:\ifnum\minute<10 0\fi\number\minute}
\def\draftlabel#1{{\@bsphack\if@filesw {\let\thepage\relax
   \xdef\@gtempa{\write\@auxout{\string
      \newlabel{#1}{{\@currentlabel}{\thepage}}}}}\@gtempa
   \if@nobreak \ifvmode\nobreak\fi\fi\fi\@esphack}
        \gdef\@eqnlabel{#1}}
\def\@eqnlabel{}
\def\@vacuum{}
\def\draftmarginnote#1{\marginpar{\raggedright\scriptsize\tt#1}}
\def\draft{\oddsidemargin -.5truein
        \def\@oddfoot{\sl preliminary draft \hfil
        \rm\thepage\hfil\sl\today\quad\militarytime}
        \let\@evenfoot\@oddfoot \overfullrule 3pt
        \let\label=\draftlabel
        \let\marginnote=\draftmarginnote
   \def\@eqnnum{(\theequation)\rlap{\kern\marginparsep\tt\@eqnlabel}%
\global\let\@eqnlabel\@vacuum}  }

\def\titlepage{\@restonecolfalse\if@twocolumn\@restonecoltrue\onecolumn
     \else \newpage \fi \thispagestyle{empty}\c@page\z@ 
        \def\thefootnote{\fnsymbol{footnote}} }

\def\endtitlepage{\if@restonecol\twocolumn \else  \fi
        \def\thefootnote{\arabic{footnote}}
        \setcounter{footnote}{0}}  

\catcode`@=12
\relax

\def\bea{\begin{array}}
\def\bem{\begin{displaymath}}
\def\beq{\begin{equation}}

\def\eea{\end{array}}
\def\eem{\end{displaymath}}
\def\eeq{\end{equation}}

\def\ket#1{\left| #1\right\rangle}

\def\ov{\overline}

\def\s2w{\sin^2 \theta_W}
\def\Tr{\mathop{\rm Tr}}

\relax
\def\crbig{\\\noalign{\vspace {3mm}}}

\newcommand{\md}{\mathrm{d}}
\def\ov{\overline}

\newcommand{\skipthispart}[1]{}

\newcommand{\bep}{\begin{picture}}
\newcommand{\eep}{\end{picture}}

\newcommand{\boldpic}[1]{{\linethickness{0.4mm}#1}}

\newcounter{YoungHeight}\newcounter{YoungWidth}

\newcounter{Mul1}\newcounter{Mul2}\newcounter{Mul3}\newcounter{Mul4}
\newcounter{A0}\newcounter{A1}\newcounter{A2}
\newcounter{B3}
\newcounter{C3}\newcounter{C4}
\newcounter{D1}\newcounter{D2}\newcounter{D3}
\newcounter{T0}\newcounter{T1}

\newlength{\txtHShift}

\newlength{\txtWidth}

\newcommand{\HalfLength}[2]{\setcounter{Mul1}{#1}\setcounter{Mul2}{#1}\addtocounter{Mul1}{\value{Mul2}}\addtocounter{Mul1}{\value{Mul2}}%
\addtocounter{Mul1}{\value{Mul2}}\addtocounter{Mul1}{\value{Mul2}}\setcounter{#2}{\value{Mul1}}}

\newcommand{\Add}[3]{\setcounter{#1}{#2}\addtocounter{#1}{#3}}

\newcommand{\Length}[1]{#10}
\newcommand{\YoungScale}{}

\newcommand{\shiftedText}[2]{{\hspace{#1}#2}}
\newcommand{\calcHShift}[1]{\settowidth{\txtWidth}{#1}\setlength{\txtHShift}{-0.5\txtWidth}}

\newcommand{\TextCenterB}[3]{{\calcHShift{#1}\HalfLength{#2}{T0}\Add{T1}{\Length{#3}}{-7}\put(\value{T0},\value{T1}){\shiftedText{\txtHShift}{#1}}}}

\newcommand{\TextTop}[3]{{\calcHShift{#1}\HalfLength{#2}{T0}\Add{T1}{\Length{#3}}{-9}\put(\value{T0},\value{T1}){\shiftedText{\txtHShift}{#1}}}}

\newcommand{\BlockA}[2]{{\YoungScale\bep(\Length{#1},\Length{#2}){\Add{A1}{#1}{1}\Add{A2}{#2}{1}}%
\multiput(0,0)(10,0){\value{A1}}{\line(0,1){\Length{#2}}}\multiput(0,0)(0,10){\value{A2}}{\line(1,0){\Length{#1}}}%
\setcounter{YoungHeight}{\Length{#2}}\setcounter{YoungWidth}{\Length{#1}}\eep}}

\newcommand{\RectT}[3]{\bep(\Length{#1},\Length{#2})\put(0,0){\line(1,0){\Length{#1}}}\put(0,0){\line(0,1){\Length{#2}}}%
\put(\Length{#1},\Length{#2}){\line(-1,0){\Length{#1}}}\put(\Length{#1},\Length{#2}){\line(0,-1){\Length{#2}}}#3{#1}{#2}\eep}

\newcommand{\RectARow}[2]{\parbox{\Length{#1}pt}{\bep(\Length{#1},10)\put(0,0){\RectT{#1}{1}{\TextTop{#2}}}\eep}}

\newcommand{\RectARowBlock}[3]{\parbox{\Length{#1}pt}{{\bep(0,0)#3\eep\Add{A0}{\value{YoungHeight}}{10}\bep(\Length{#1},\value{A0})%
\put(0,\value{YoungHeight}){\RectT{#1}{1}{\TextTop{#2}}}\eep}}}

\newcommand{\RectBRow}[4]{\parbox{\Length{#1}pt}{\bep(\Length{#1},20)\put(0,0){\RectT{#2}{1}{\TextTop{#4}}}%
\put(0,10){\RectT{#1}{1}{\TextTop{#3}}}\eep}}

\newcommand{\RectCRow}[6]{\parbox{\Length{#1}pt}{\bep(\Length{#1},30)\put(0,0){\RectT{#3}{1}{\TextTop{#6}}}%
\put(0,10){\RectT{#2}{1}{\TextTop{#5}}}\put(0,20){\RectT{#1}{1}{\TextTop{#4}}}\eep}}

\newcommand{\TextTopA}[3]{{\calcHShift{#1}\HalfLength{#2}{T0}\Add{T1}{\Length{#3}}{-7}\put(\value{T0},\value{T1}){\shiftedText{\txtHShift}{#1}}}}
\newcommand{\RectDRow}[8]{\parbox{\Length{#1}pt}{\bep(\Length{#1},40)\put(0,0){\RectT{#4}{1}{\TextTopA{#8}}}%
\put(0,10){\RectT{#3}{1}{\TextTopA{#7}}}%
\put(0,20){\RectT{#2}{1}{\TextTopA{#6}}}\put(0,30){\RectT{#1}{1}{\TextTopA{#5}}}\eep}}

\newcommand{\RectBRowUp}[4]{\parbox{\Length{#1}pt}{\bep(\Length{#1},20)\put(0,0){\RectT{#2}{1}{\TextCenterB{#4}}}%
\put(0,10){\RectT{#1}{1}{\TextCenterB{#3}}}\eep}}

\newcommand{\YoungC}{\BlockA{3}{1}}

\newcommand{\BlockApar}[2]{\parbox{\Length{#1}pt}{\YoungScale\bep(\Length{#1},\Length{#2}){\Add{A1}{#1}{1}\Add{A2}{#2}{1}}%
\multiput(0,0)(10,0){\value{A1}}{\line(0,1){\Length{#2}}}\multiput(0,0)(0,10){\value{A2}}{\line(1,0){\Length{#1}}}%
\setcounter{YoungHeight}{\Length{#2}}\setcounter{YoungWidth}{\Length{#1}}\eep}}

\newcommand{\BlockBpar}[4]{\parbox{\Length{#1}pt}{\YoungScale\Add{B3}{\Length{#2}}{\Length{#4}}%
\bep(\Length{#1},\value{B3})\put(0,\Length{#4}){\BlockA{#1}{#2}}%
\put(0,0){\BlockA{#3}{#4}}\setcounter{YoungHeight}{\value{B3}}\setcounter{YoungWidth}{\Length{#1}}\eep}}

\newcommand{\YoungpA}{\BlockApar{1}{1}}
\newcommand{\YoungpB}{\BlockApar{2}{1}}
\newcommand{\YoungpC}{\BlockApar{3}{1}}
\newcommand{\YoungpD}{\BlockApar{4}{1}}
\newcommand{\YoungpE}{\BlockApar{5}{1}}
\newcommand{\YoungpF}{\BlockApar{6}{1}}

\newcommand{\YoungpAA}{\BlockApar{1}{2}}
\newcommand{\YoungpBA}{\BlockBpar{2}{1}{1}{1}}
\newcommand{\YoungpBB}{\BlockApar{2}{2}}
\newcommand{\YoungpCA}{\BlockBpar{3}{1}{1}{1}}

\newcommand{\YoungpDA}{\BlockBpar{4}{1}{1}{1}}

\newcommand{\YoungpAAA}{\BlockApar{1}{3}}

\newcommand{\pl}{\partial}
\usepackage[numbers,sort&compress]{natbib}
\usepackage[nottoc]{tocbibind}

\newcommand{\fud}[2]{{}^{#1}{}_{#2}\,}

\newcommand{\YY}[1]{\boldsymbol{Y}(#1)}
\newcommand{\YYY}{\boldsymbol{Y}}

\relax

\begin{document}
\pagenumbering{gobble}
\hfill
\vskip 0.01\textheight
\begin{center}
{\Large\bfseries 
An excursion into the string spectrum}

\vspace{0.4cm}

\vskip 0.03\textheight
\renewcommand{\thefootnote}{\fnsymbol{footnote}}
Chrysoula \textsc{Markou} and 
Evgeny \textsc{Skvortsov}\footnote{Research Associate of the Fund for Scientific Research -- FNRS, Belgium.}\footnote{Also on leave from Lebedev Institute of Physics.}
\renewcommand{\thefootnote}{\arabic{footnote}}
\vskip 0.03\textheight

{\em Service de Physique de l'Univers, Champs et Gravitation, \\ Universit\'e de Mons, 20 place du Parc, 7000 Mons, 
Belgium}\\

\end{center}

\vskip 0.02\textheight

\begin{abstract}
We propose a covariant technique to excavate physical bosonic string states by entire trajectories rather than individually. The approach is based on Howe duality: the string's spacetime Lorentz algebra commutes with a certain inductive limit of $sp(\bullet)$, with the Virasoro constraints forming a subalgebra of the Howe dual algebra $sp(\bullet)$. There are then infinitely many simple trajectories of states, which are lowest--weight representations of $sp(\bullet)$ and hence of the Virasoro algebra. Deeper trajectories are recurrences of the simple ones and can be probed by suitable trajectory--shifting operators built out of the Howe dual algebra generators. We illustrate the formalism with a number of subleading trajectories and compute a sample of tree--level amplitudes.
\end{abstract}

\newpage
\tableofcontents
\newpage
\section{Introduction}
\label{sec:}
\pagenumbering{arabic}
\setcounter{page}{2}

\fancyhead{} 
\fancyfoot{} 
\fancyfoot[C]{\thepage\ of \pageref{LastPage}}
 
\renewcommand{\headrulewidth}{0pt} 

The string spectrum is crowded by massive higher--spin and, more generally, mixed--symmetry states, all of which are vital for the theory's  consistency. While most of its extent remains little explored, some examples of level--by--level probes of a few of the lightest string states include \cite{Koh:1987hm, Feng:2010yx, Bianchi:2010es, Feng:2011qc, Feng:2012bb, Lust:2021jps, Benakli:2021jxs, Benakli:2022ofz, Benakli:2022edf, Lust:2023sfk} and, while the leading Regge trajectory is well understood, see e.g.  \cite{Weinberg:1985tv,Manes:1988gz,Sagnotti:2010at, Schlotterer:2010kk, Bianchi:2011se, Cangemi:2022abk}, much less is known about the subleading ones, see e.g. \cite{Weinberg:1985tv,Manes:1988gz}. Various other aspects have been investigated, among which vertex operators for coherent string states \cite{Hindmarsh:2010if, Skliros:2011si}, the decay of superheavy string states \cite{Feng:2012bb, Gross:2021gsj}, with evidence for chaotic traits having recently been presented \cite{Gross:2021gsj, Rosenhaus:2021xhm, Firrotta:2022cku, Bianchi:2022mhs, Bianchi:2023uby}, as well as the renormalization of the mass spectrum \cite{Pius:2013sca, Pius:2014iaa, Sen:2016gqt}. In addition, in the context of twisted strings \cite{Hohm:2013jaa, Huang:2016bdd,Jusinskas:2021bdj}, a new approach to the computation of tree–level string amplitudes of one massive and arbitrarily many massless string states, employing both string and field theory tools, has appeared \cite{Guillen:2021mwp}. Yet another line of research that is sensitive to the presence of higher--spin states is the exploration of the universality of Veneziano(--like) amplitudes and of Regge trajectories \cite{Caron-Huot:2016icg,Sever:2017ylk}.

There is a number of well--known techniques that allow one, as a matter of principle, to get access to any string state. Firstly, in the light--cone gauge \cite{Goddard:1972iy, Goddard:1973qh} any function of the transverse oscillators $\alpha^i_{-n}$, $i=1,\ldots, d-2$, represents a physical state. Secondly, the DDF formalism \cite{DelGiudice:1971yjh,Brower:1972wj} allows one to climb up the string spectrum by scattering photons multiple times off the lowest state. While the two approaches do cover the entirety of string spectrum, in both cases some external reference momenta are needed to fix the frame, which is intrinsically non--covariant. In addition, one does not easily reach elementary states, obtaining instead by default a linear superposition thereof or even fractions of elementary states. The smallest or elementary string states can be thought of as corresponding to particles in the sense of the Wigner classification \cite{Wigner:1939cj}, see \cite{Bekaert:2006py} for a review and for its $d$--dimensional generalizations that are relevant for string theory. Physical string states namely correspond to irreducible representations of Wigner's little algebra, which is $so(d-1)$ for massive particles and $so(d-2)$ for massless ones. 

For the purpose of exploring the string spectrum and navigating therein, it would be advantageous to have at our disposal a technique that gives easy access to elementary string states instead of superpositions thereof, which is precisely the goal of this paper for the case of the bosonic string. Our simple technical observation is that there exists a relevant algebra that is bigger than the Virasoro one: it is the inductive limit of $sp(\bullet)$ that operates on the mode labels $n$ of $\alpha_{-n}^\mu$ or $\pl^n X^\mu$. Any irreducible state activates a finite number of creation oscillators, say with $n\in[0,K]$ at a $K$ bounded by its mass level, in which case the relevant algebra becomes $sp(2K)$. The latter commutes with the spacetime Lorentz algebra and this pair can be recognized as \textit{Howe duals} \cite{Howe}. When represented on such states, the Virasoro constraints reduce to a subalgebra of $sp(2K)$. The simplest physical states are then the lowest--weight states of the full $sp(2K)$ and, hence, of the Virasoro algebra. It is important to note that confining ourselves to $K$ first oscillators does not restrict the level and the $sp(2K)$ lowest--weight condition covers \textit{entire} trajectories. Therefore, the simplest trajectories are built from the lowest--weight states of $sp(2K)$. Other trajectories are recurrences of the simplest ones and can be reached with the help of dressing functions, namely \textit{trajectory--shifting} operators that are built out of creation operators of $sp(2K)$. 

An advantage of our approach is that it covers infinitely many states of the ``same complexity'': if we manage to reach a single physical state belonging to a given trajectory by constructing a suitable dressing function, the \textit{same} function covers \textit{all} similar states of the entire trajectory. By ``similar states'' we refer to the states that have the same type of Young diagram specifying the respective $so(D-1)$ representation, e.g. all symmetric polarization tensors or all polarization tensors with the symmetry of Young diagrams with $n$ rows. The complexity of dressing functions depends on two main parameters: i) the ``\textit{depth}'', which we can think of as the difference of levels between the lowest--level appearance of a given physical Young diagram to the higher level it can also appear and which we want to get access to; ii) the \textit{complexity} of the polarization tensors of the corresponding physical states, which, in practice, is the number of rows of the Young diagram.

While the DDF and the light--cone gauge are two very general approaches in that they offer free access to the whole string spectrum, there is a number of techniques that require additional calculations to yield an actual state: (i.a) one can try to solve the Virasoro constraints directly by decomposing all states at a given level into irreducible ones, which is tedious and gives access to a few light levels, see e.g. \cite{Sasaki:1985py,Tanii:1986ug}; (i.b) a  more economical way is to consider tensors that are transverse with respect to the momentum \cite{Manes:1988gz}; (ii) the vertex operators can also be fixed by starting with their covariant versions and requiring local Weyl invariance \cite{Weinberg:1985tv,Polchinski:1986qf,Ichinose:1986vj,DHoker:1987rxo}, which is rather cumbersome to apply. Some of the papers that address infinitely many irreducible states include \cite{Weinberg:1985tv,Manes:1988gz,Bianchi:2010es,Sagnotti:2010at,Schlotterer:2010kk,Skliros:2011si,Gross:2021gsj,Bianchi:2023uby}. If one is interested in the spectrum only rather than vertex operators, the character(ization) has been worked out in \cite{Hanany:2010da}, see also \cite{Lust:2012zv}. Our approach is somewhat in between: it covers infinitely many states at low cost, including many of the subleading ones; probing trajectories beyond the simplest also requires further calculations which, nevertheless, always excavate states by \textit{entire} trajectories. 

The paper is organized as follows. In section \ref{sec:ingredients} we provide a short overview of bosonic string theory, which mainly recalls the standard notation and normalizations. We also review there what is known about the string spectrum and give examples of vertex operators. In section \ref{sec:excavation} we develop the technique of excavating trajectories with the help of dressing functions and Howe duality. Many examples are given in section \ref{sec:slices}. To illustrate the formalism some amplitudes are computed in section \ref{sec:amplitudes}. Conclusions and discussion can be found in section \ref{sec:conclusions}. There are also three Appendices with small technical details referred to in due time.

\textbf{Conventions.} We use the mostly plus metric signature, so that the mass $m$ and the momentum $p^\mu$ of a state at rest are related via $m^2=-p^2\,$. We assume a flat background spacetime and use Greek letters $\mu,\nu, \ldots$ to denote indices of the Lorentz algebra $so(d-1,1)$. Whenever there is a shortage of letters, we also use Latin letters $a,b, \ldots$ for the same purpose, which are also used for abstract tensors. All vertex operators and energy--momentum tensors are implicitly normal--ordered throughout this work.

\section{Bosonic string ingredients}
\label{sec:ingredients}


\subsection{Worldsheet CFT and the operator--state correspondence}
\label{subsec:fundams}

In this section we review the rudiments (see for example \cite{Polchinski:1998rq,Blumenhagen:2013fgp}) of bosonic string theory in regard to its worldsheet properties and physical spectrum, on which we will build in later sections. The string spectrum is essentially a property of worldsheet actions; the bosonic string in particular is described by the Polyakov action for the \textit{real} field $X^\mu(z,\ov{z})$, that can be written as
\begin{align} \label{Polyakov}
    S_{\textrm{P}}=\frac{1}{2\pi \alpha'} \int \md^2 z \, \partial X \cdot \ov{\partial} X \quad , \quad T =\frac{1}{2\pi \alpha'}\,,
\end{align}
where $\alpha'$ is the string scale, related to the string tension $T$ as indicated above. $X^\mu$ is a map from a $2$--dimensional worldsheet, that we parametrize directly in complex coordinates  $(z,\ov{z})$, to a $26$--dimensional spacetime, in which the (Minkowski) scalar product we denote by ``$\cdot$''; $\mu,\nu,\ldots=0, \ldots, 25\,$ denote spacetime Lorentz indices, but in most of this work we keep their range free, namely $\mu,\nu,\ldots=0,\ldots,d-1$. The form (\ref{Polyakov}) implies a conformal gauge--fixing of the worldsheet metric and a Euclidean worldsheet signature. The equation of motion derived from (\ref{Polyakov}) reads
\begin{align}
     \partial \ov{\partial}X^\mu(z,\ov{z})=0 \,,
\end{align}
which for closed strings implies the Fourier expansion
\begin{align}
 \label{closed_expansion}
X^\mu(z,\ov{z}) = x^\mu - i \frac{\alpha'}{2} k^\mu \ln |z|^2 + i \sqrt{\frac{\alpha'}{2}} \sum_{n \neq 0} \frac{1}{n} \bigg[ \frac{\alpha_n^\mu}{z^n} + \frac{\ov{\alpha}_n^\mu}{\ov{z}^n} \bigg] \,,
\end{align}
while for open strings with Neumann boundary conditions 
\begin{align}
 \label{open_expansion}
X^\mu(z,\ov{z}) = x^\mu - i \alpha' p^\mu \ln |z|^2 + i \sqrt{\frac{\alpha'}{2}} \sum_{n \neq 0 } \frac{\alpha_n^\mu}{n} \bigg[ \frac{1}{z^n} +  \frac{1}{\ov{z}^n} \bigg]\,, 
\end{align}
where $x^\mu$ and ($k^\mu$) $p^\mu$ are respectively the position and momentum of the (closed) open string's center of mass. 

The open string expansion (\ref{open_expansion}) can be obtained from the closed string one (\ref{closed_expansion}) by setting
\begin{align}
    k^\mu = 2p^\mu \quad , \quad \alpha_n^\mu = \ov{\alpha}_n^\mu\,,
\end{align}
since a closed string can be thought of as two open strings with identified ends, each carrying half its momentum; the open string's boundary conditions also force the presence of a single set of Fourier modes $\alpha_n^\mu$. It is instructive to further set
\begin{align}
      \alpha^\mu_0 \equiv \begin{cases}
      \sqrt{\frac{\alpha'}{2}} \, k^\mu \quad , \quad \textrm{for closed strings}
      \crbig
          \sqrt{2\alpha'} \, p^\mu \quad , \quad \textrm{for open strings}\,.
      \end{cases} 
\end{align}
The worldsheet topology of closed and open strings can be thought of as that of the (Riemann) sphere $S^2$ and the disk $\mathbb{D}_2$ respectively, the boundary of the latter being the real axis $\mathbb{R}$. Open string relations can then be obtained from their closed string versions by means of the doubling trick \cite{Cardy:1989ir}, which can be formulated as the substitutions
\begin{align} \label{doubling}
    X \rightarrow \frac{1}{2} X \quad \textrm{or} \quad \alpha' \rightarrow 4\alpha'\,.
\end{align}

The $\alpha^\mu_n$, that will be referred to as the (bosonic) oscillators in the following as is customary, obey the reality condition and the commutation relation 
\begin{align}
    (\alpha_n^\mu)^\dagger = \alpha_{-n}^\mu \quad , \quad    [\alpha^\mu_m,\alpha^\nu_n]=m \,\delta_{m+n,0} \,\eta^{\mu\nu}
\end{align}
and the $\ov{\alpha}_n^\mu$ obey another copy of the same algebra in the case of closed strings. Consequently, for $n>0$, $\alpha_n^\mu$ and $\alpha_{-n}^\mu $ act up to $\sqrt{n}$ as annihilation and creation operators respectively, namely
\begin{align}
    \alpha_n^\mu \ket{0;p} = 0 \quad , \quad n>0
\end{align}
for the vacuum state $\ket{0;p}\,$. A number operator can be defined according to
\begin{align} \label{numberop}
    N_n \equiv \begin{cases}
         \frac{1}{n} \, :\alpha_n \cdot \alpha_{-n}:\, = \frac{1}{n}\, \alpha_{-n} \cdot \alpha_n \quad , \quad n>0
         \crbig
         0\quad , \quad n=0
    \end{cases} \quad , \quad N \equiv \sum_{n=0} n N_n \,.
\end{align}

A generic open string state of momentum $p^\mu$ can then be represented as
\begin{align} \label{gen_op_st}
    \ket{\phi} =\phi_{\mu_1 \ldots \mu_k} (p) \, \alpha^{\mu_1}_{-n_1}  \ldots \alpha^{\mu_k}_{-n_k} |0;p \rangle   \quad , \quad N =\displaystyle \sum_{i=1}^k n_i \quad , \quad n_i >0
\end{align}
where $\phi_{\mu_1 \ldots \mu_k} (p)$ is an a priori arbitrary tensor which renders $\ket{\phi}$ a spacetime scalar and each oscillator may appear with a different occupation number. In the old covariant quantization, $\ket{\phi}$ is physical provided it satisfies the conditions\footnote{In the familiar form these constraints were written first in \cite{DelGiudice:1970dr}.}
\begin{align} \label{GBphys}
     (L_n -\,\delta_{n,0})\ket{\phi}= 0 \quad , \quad \forall \, n \in \mathbb{N}\,,
\end{align}
where the Virasoro operators $L_n$ are given by
\begin{align}
    L_n=  \tfrac12 \sum_{m=-\infty}^{+\infty} \, :\alpha_{n-m}\cdot \alpha_m : \,.
\end{align}
They satisfy the famous Virasoro algebra
\begin{align} \label{vir_alg}
    [L_n,L_k]&=(n-k)L_{n+k}+\frac{c}{12}(n^3-n)\delta_{n+k,0}\,.
\end{align}
Since the subalgebra of ``positive'' operators $L_n$, $n>0\,$, is generated by $L_1$ and $L_2$, it is \textit{sufficient} to impose
\begin{align}
     L_1 \ket{\phi}= 0  \quad , \quad L_2 \ket{\phi}= 0 \,,
\end{align}
see for example \cite{Sasaki:1985py}, along with the lowest Virasoro constraint $(L_0 -1)\ket{\phi}= 0$, which provides the mass spectrum 
\begin{align} \label{open_mass}
    M^2_{\textrm{o}}= \frac{N-1}{\alpha'} \,,
\end{align}
with $N$ referred to as the mass level. For closed string states, two copies of (\ref{gen_op_st}) are needed, one for the left and one for the right movers, with level--matching yielding
\begin{align} \label{closed_mass}
    M^2_{\textrm{c}}= (N^L-1) \frac{4}{\alpha'} =(N^R-1) \frac{4}{\alpha'}  \,.
\end{align}

Differentiating (\ref{closed_expansion}) and inverting gives 
\begin{align}\label{dictionary}
     \partial X^\mu (z) =-i \sqrt{\frac{\alpha'}{2}} \, \sum_{n=-\infty}^{+\infty} \frac{\alpha_{n}^\mu}{z^{n+1}} \quad \Rightarrow \quad \alpha^\mu_{-n} \sim  \frac{i}{\sqrt{2\alpha'}}   \, \frac{1}{(n-1)!} \, \partial^n X^\mu \quad, \quad n>0\,,
\end{align}
where the doubling trick has been employed; the (open string) creation operators are namely the Laurent coefficients of the regular part of $\partial X^\mu(z)$. The dictionary (\ref{dictionary}) provides a ``1--1'' correspondence between oscillators and fields, or more generally between states and vertex operators \cite{Sasaki:1985py}. In particular, an open string state is said to be created by a vertex operator $V(z)$ that is a conformal field within the worldsheet conformal field theory, the ingredients of which are the ``matter'' fields $X^\mu(z)$ and the (anticommuting) conformal ghost system $b(z), c(z)$. They are described by the worldsheet energy--momentum tensors
\begin{align}\label{EM}
T(z)= -\frac{1}{4\alpha'}  \partial X \cdot \partial X\,(z) \quad , \quad T^{b,c}(z) = -2\,b\,\partial c\,(z)- (\partial b)\,c\,(z)\,,
\end{align}
with respect to which $\partial X^\mu(z)$ is a conformal primary of weight $1$ and $\partial^k X^\mu (z)$ a descendant of weight $k$. The relevant 2-- and 3--point functions read
\begin{align} \label{boson_corr}
\langle \pl X^\mu(z_1) \pl X^\nu(z_2) \rangle_{\mathbb{D}_2} = - 2 \alpha' \eta^{\mu \nu} \, \frac{1}{z_{12}^2}
\end{align}
and
\begin{align} 
 \langle c(z_1)b(z_2) \rangle = \frac{1}{z_{12}} \quad , \quad  \langle  c(z_1)c(z_2)c(z_3) \rangle = z_{12}z_{13}z_{23} \,,
\end{align}
where $z_{ij} \equiv z_i - z_j$ and the worldsheet BRST charge is  given by
\begin{align} \label{charge_b}
    Q=\oint \frac{\md z}{2\pi i}\Big[c \big(T+\frac{1}{2}T^{b,c}\big) \Big] =\oint \frac{\md z}{2\pi i} \,c \, \Big[T+(\partial c) b   \Big]\,.
\end{align}
A generic open string \textit{integrated} vertex operator $\mathcal{V}_F$ of weight $h$ can then be written as
\begin{align} \label{generic_vo_polynomial}
\mathcal{V}_F=\int \md z\, V_F(z)\quad , \quad  V_F(z) = F\Big(\partial X^\mu(z), \partial^2 X^\mu(z), \ldots, \partial^k X^\mu(z)\Big) \, e^{ip\cdot X(z)}\,,
\end{align}
where $F$ is an a priori arbitrary polynomial of $\pl X$ and its descendants and $e^{ip\cdot X}$ the momentum eigenstate of weight $\alpha'p^2$. For normalization purposes, each $\partial X$ or descendant should appear together with a prefactor of $\frac{i}{\sqrt{2\alpha'}}\,$, which cancels its mass dimension. The physical state condition in the BRST quantization takes the form
\begin{align}    \label{gen_phys}
[Q,V_F] = \textrm{tot. deriv.} \quad \Rightarrow \quad [Q,V_F] = \partial (cV_F) \quad \textrm{or} \quad h = 1 \,,
\end{align}
namely for all physical vertex operators 
\begin{align}  \label{ope_general_physical}
T(z) \, V_F(w) \,\sim \, \frac{V_F(w)}{(z-w)^2} + \frac{\partial V_F(w)}{z-w}\,.
\end{align}
At level $N$, the weights $h_{\textrm{eigen}}$ and $h_F$ of $e^{ip\cdot X}$ and $F$ are respectively given by
\begin{align} \label{weights}
    h_{\textrm{eigen}} =1-N \quad , \quad h_F = N
\end{align}
due to (\ref{open_mass}) and (\ref{gen_phys}). For example, the vertex operator of a symmetric rank--$N$ tensor at level $N$, the set of which defines the leading Regge trajectory, reads
\begin{align}  \label{vo_gen}
V_{\phi}(p,z) = \bigg(\frac{i}{\sqrt{2\alpha'}}\bigg)^N \, \phi_{\mu_1 \ldots \mu_N} (p) \, \partial X^{\mu_1}(z) \ldots \partial X^{\mu_N}(z) \, e^{ip \cdot X(z)}\,.
\end{align}
This Ansatz obviously satisfies (\ref{weights}) and the on--shell condition  (\ref{open_mass}); absence of higher--order poles in (\ref{ope_general_physical}) further enforces transversality and tracelessness
\begin{align}
 p^\mu \phi_{\mu \mu_2 \ldots \mu_N} = 0 \quad , \quad \phi^{\mu}_{\hphantom{\mu} \mu \mu_3 \ldots \mu_N} =0\,.
\end{align}

Let us also note that (\ref{gen_phys}) is equivalent to imposing
\begin{align} \label{phys_unint}
    [Q,V]=0
\end{align}
on the \textit{unintegrated} operator
\begin{align} \label{unintegrated}
    V(z)=c(z)\,V_F(z)\,.
\end{align}
Moreover, any operator of the form $[Q,U]$, where $U$ is \textit{any} operator, satisfies (\ref{phys_unint}) due to the Jacobi identity. Consequently, BRST--exact states can be represented by vertex operators of the form (\ref{unintegrated}), so that we can find such spurious vertex operators $V_{\textrm{sp}}$ via
\begin{align}\label{spurious_def}
c V_{\textrm{sp}}= [Q,U] \quad \Rightarrow \quad  V_{\textrm{sp}}= \partial U\,,
\end{align}
where $V_{\textrm{sp}}$ still has to satisfy (\ref{gen_phys}), so $U$ must have weight $0$, equivalently $\partial U$ weight $1$. Eq. \eqref{spurious_def} is equivalent to the action of $L_{-1}$ in the old covariant quantization. There is one more generator of null states, $L_{-2}$, whose action can be obtained via $\delta V= [Q, b U]$ for an integrated operator $V$.  

As a final comment, a factor of $g_{\textrm{o}} T^a$, where $g_{\textrm{o}}$ is the (dimensionful) open string coupling and $T^a$ the brane group generator à la Chan--Paton, is assumed to dress every physical open string vertex operator; $g_{\textrm{o}}$ essentially sets the strength of the states' interactions and eventually ensures the correct mass dimension of scattering amplitudes. Physical closed string vertex operators are obtained by taking two copies of (\ref{generic_vo_polynomial}), one for the left-- and one for the right--movers, and imposing the level--matching condition (\ref{closed_mass}), so they have a holomorphic and an antiholomorphic part with weights $(1,1)$. Each closed string vertex operator is then dressed by the closed string coupling $g_{\textrm{c}} \sim g_{\textrm{o}}^2$.

\subsection{Examples of light vertex operators}
\label{sub:lightex}

One way of constructing physical string states is by choosing a level $N$, which determines the weight (\ref{weights}) of the polynomial $F$ of the level's vertex operator (\ref{generic_vo_polynomial}). One then proceeds to write an Ansatz for $F$, where a priori arbitrary Lorentz tensors contract $\partial X^\mu$ and its descendants to formulate spacetime scalars. Computing the OPE with $T(z)$ may yield poles of order higher than $2$ because of Wick's theorem, to cancel which and maintain (\ref{ope_general_physical}), constraints involving the various Lorentz tensors and momentum emerge. Eventually, this results in the level splitting into transverse and traceless irreducible Lorentz representations, namely on--shell physical string states. In this subsection we review this procedure for the open bosonic string up to $N=3$.

\paragraph{$\boldsymbol{N=0}$.} (\ref{weights}) yields $h_F=0$, so there is a single possible term contributing to the level's vertex operator: 
\begin{align}\label{open_tachyon}
V^{\textrm{open}}_{\textrm{tachyon}} (p,z) = e^{ip\cdot X(z)} \quad , \quad p^2 = \frac{1}{\alpha'}\,.
\end{align}
The vertex operator (\ref{open_tachyon}) creates a single state that is a tachyonic scalar and (\ref{ope_general_physical}) is automatically satisfied. 

\paragraph{$\boldsymbol{N=1}$.}  (\ref{weights}) yields $h_{F}=1,$ so again there is a single possible term contributing to the level's vertex operator: 
\begin{align} \label{open_vector}
V_{\epsilon} (p,z) = \frac{1}{\sqrt{2\alpha'}} \, \epsilon_\mu (p) \, i\partial X^\mu(z) \, e^{ip\cdot X(z)} \quad , \quad p^2=0\,,
\end{align}
where $\epsilon^\mu$ is a vector of $SO(26)$. The OPE of (\ref{open_vector}) with the energy--momentum tensor (\ref{EM}) produces, however, a pole of cubic order, to cancel which the transversality condition
\begin{align}
\epsilon \cdot p =0\,,
\end{align}
which can be thought of as removing $1$ of the generic vector's $26$ degrees of freedom, has to be enforced. It is furthermore instructive to notice how the gauge invariance 
\begin{align}
    \epsilon^\mu \rightarrow \epsilon^\mu + p^\mu\,,
\end{align}
that removes another degree of freedom, is realized. To this end, for a polarization equal to the state's momentum, the vertex operator (\ref{open_vector}) becomes
\begin{align} \label{vector_inv}
 V_{\textrm{sp}} (p,z) = \frac{1}{\sqrt{2\alpha'}} \, \partial \big( e^{ip\cdot X(z)}\big)\,,
\end{align}
namely it is a total derivative and so does not contribute to scattering amplitudes. Formally, (\ref{open_vector}) is BRST--closed and (\ref{vector_inv}) is BRST--exact, since the latter can be written as
\begin{align} \label{vector_inv_comm}
  c \,  V_{\textrm{sp}}= [Q,U ] \quad , \quad U \equiv \frac{1}{\sqrt{2\alpha'}}e^{ip\cdot X} \quad , \quad p^2=0\,,
\end{align}
which confirms that it is a spurious state corresponding to one pure gauge degree of freedom. Notice that this spurious state is associated with $U$, which resembles the tachyon vertex operator (\ref{open_tachyon}) at one level higher. The vertex operator (\ref{open_vector}) thus creates a single state that is a massless vector and propagates $24$ degrees of freedom.

\paragraph{$\boldsymbol{N=2}$.} The first treatment of this level appeared in \cite{Friedan:1985ge}. (\ref{weights}) yields $h_F=2$ so there are two possible terms\footnote{their relative unphysical prefactor of $2$ is chosen for convenience}
\begin{align}\label{open_2}
V_B(p,z)=\bigg[\frac{1}{4\alpha'} B_{\mu \nu}(p)\, i\partial X^{\mu}(z) \,i \partial X^{\nu}(z) + \frac{1}{\sqrt{2\alpha'}} \phi_\mu(p) \,i\partial^2 X^\mu (z) \bigg] e^{ip \cdot X(z)}\,,
\end{align}
where $B^{\mu \nu}$ and $\phi^\mu$ are a (by construction) symmetric rank--$2$ tensor and a vector of $SO(26)$ respectively. Since
\begin{align} \label{pione}
   \phi_\mu\, i\partial^2 X^\mu  e^{ip \cdot X} = \textrm{tot. deriv.} - \phi_\mu p_\nu\, i\partial X^{\mu} \,i \partial X^{\nu} e^{ip \cdot X}\,,
\end{align}
we already expect that (\ref{open_2}) propagates a single physical state associated with $B_{\mu \nu}$. More concretely, the OPE of (\ref{open_2}) with the energy--momentum tensor (\ref{EM}) produces a pole of quartic and one of cubic order, to cancel which one has to impose respectively the conditions
\begin{align}
    B_\mu^{\hphantom{\mu} \mu} +4\sqrt{2\alpha'} p_\mu \phi^\mu &=0 \,, \label{cond_2_1}
    \crbig
    \sqrt{2\alpha'}\, p^\mu B_{\mu \nu}+4\,\phi_\nu &=0\,.  \label{cond_2_2}
\end{align}
Interestingly, these are invariant under the \textit{restricted} Stueckelberg transformation \cite{Callan:1986ja, Labastida:1988wi}
\begin{align}
    \delta B_{\mu \nu} = 2\sqrt{2\alpha'} \, ( p_\mu \xi_\nu + p_\nu \xi_\mu) \quad, \quad \delta \phi_\mu = \xi_\mu \quad , \quad p \cdot \xi=0\,,
\end{align}
which can be used to fix the $p$--transverse components of $\phi$ to zero, $\phi^\mu_\perp=0\,$, which further implies that (\ref{cond_2_1}) and (\ref{cond_2_2}) become tracelessness and transversality conditions
\begin{align} \label{cond_massive_spin_2}
B_\mu^{\hphantom{\mu} \mu} =0 \quad , \quad p^\mu B_{\mu \nu} =0 
\end{align}
and
\begin{align}
    p^\nu \phi_\nu=0\,.
\end{align}

This gauge--fixing amounts to the $25$ d.o.f. of the spurious massive vector
\begin{align} \label{spurious_2}
 V_{\textrm{sp}, \xi} (p, z) = \frac{1}{\sqrt{2\alpha'}} \, \partial \big( \xi_\mu \,i \partial X^\mu(z) \, e^{ip\cdot X(z)}\big) \quad , \quad p\cdot \xi=0 \quad , \quad p^2=-1/\alpha'
\end{align}
for which
\begin{align} \label{spurious_2_comm}
   c \,  V_{\textrm{sp}, \xi} (p, z) =  [Q,U_\xi]  \quad , \quad U_\xi \equiv \frac{1}{\sqrt{2\alpha'}} \xi_\mu \, i \partial X^\mu \,e^{ip\cdot X}  \,.
\end{align}
Consequently, there appears a single physical state at level $N=2\,$,
\begin{align} \label{lightestgr}
    V_B(p,z) = \frac{1}{2\alpha'} \, B_{\mu \nu} (p) \, i\partial X^{\mu}(z)  i\partial X^{\nu}(z) \, e^{ip \cdot X(z)}\,,
\end{align}
subject to (\ref{cond_massive_spin_2}), propagating $324$ d.o.f.

\paragraph{$\boldsymbol{N=3}$.} (\ref{weights}) yields $h_F=3$ so there are three possible terms
\begin{align} \label{open_3}
    V_{N=3} = \bigg[ \frac{1}{(2\alpha')^{3/2}} \, F_{\mu \nu \lambda} \, i\partial X^{\mu}  i\partial X^{\nu}  i\partial X^{\lambda} + \frac{1}{2\alpha'} F_{\mu \nu }  \, i\partial^2 X^{\mu}  i\partial X^{\nu} + \frac{1}{\sqrt{2\alpha'}}F_\mu \, i \partial^3 X^\mu \bigg] \, e^{ip \cdot X} \,,
\end{align}
where $F_{\mu \nu \lambda}(p)$, $F_{\mu \nu}(p)$ and $F_\mu(p)$ are a (by construction) fully symmetric rank--$3$ tensor, a rank--$2$ tensor and a vector of $SO(26)$ respectively. Since
\begin{align} \label{pitwo}
    F_\mu \, i\partial^3 X^\mu\, e^{ip \cdot X} &= \textrm{tot. deriv.} - F_\mu p_\nu \, i\partial^2 X^\mu  i\partial X^\nu \, e^{ip \cdot X}\,,
\crbig
 2 F_{(\mu \nu)} \, i\partial^2 X^{\mu}  i\partial X^{\nu} \, e^{ip \cdot X} &=  \textrm{tot. deriv.} -F_{\mu \nu} p_\lambda \, i\partial X^\mu \, i\partial X^\nu \partial X^\lambda \, e^{ip \cdot X}\,,
\end{align}
 we already expect that $F_{[\mu \nu]}$ and $F_{\mu \nu \lambda}$ are sufficient to determine the physical content of (\ref{open_3}). More concretly, the OPE of (\ref{open_3}) with the energy--momentum tensor (\ref{EM}) produces poles of order $5$, $4$, $3$, cancelation of which yields
\begin{align} \label{gauge_N2}
 F_\mu^{\hphantom{\mu} \mu} =0\,,
 \crbig
    3F_{\mu\nu\lambda} \eta^{\mu \nu}+2\sqrt{2\alpha'}p^\mu F_{\mu \lambda} +6F_\lambda =0\,, 
    \crbig
   p^\mu F_{\mu \nu \lambda} =0 \quad , \quad  \sqrt{2\alpha'}p^\mu F_{\mu \lambda} +6F_\lambda =0 \quad , \quad p^\mu F_\mu =0  \,,
\end{align}
using which one can eliminate $F_\mu$ and $F_{(\mu \nu)}$. There are then two physical states at this level \cite{Weinberg:1985tv, Sasaki:1985py}, namely the symmetric rank--$3$ tensor
\beq
\begin{array}{ccl}
V_{F^{\mu \nu \lambda}}(p,z) = \frac{1}{(2\alpha')^{3/2}}  \, F_{\mu \nu \lambda} (p) \, i\partial X^{\mu}(z)  i\partial X^{\nu}(z)  i\partial X^{\lambda}(z) \, e^{ip \cdot X(z)}
\crbig
p^2 = -\frac{2}{\alpha'} \quad , \quad p^\mu F_{ \mu \nu \lambda} = 0 \quad , \quad F^{\mu}_{\hphantom{\mu} \mu \nu } =0
\end{array}
\eeq
as well as the antisymmetric rank--$2$ tensor
\beq
\begin{array}{ccl}
V_{F^{\mu \nu }}(p,z) = \frac{1}{2\alpha'}\, F_{\mu \nu } (p) \, i\partial^2 X^{\mu}  i\partial X^{\nu} (z) \, e^{ip \cdot X(z)}
\crbig
p^2 = -\frac{2}{\alpha'} \quad , \quad p^\mu F_{ \mu \nu} = 0 \quad , \quad F_{(\mu \nu) }=0\,.
\end{array}
\eeq

By inspecting the partial integrations e.g. (\ref{pione}) and (\ref{pitwo}) associated with the spurious states at every level, we can expect more generally that terms \textit{linear} in $p^\mu$ within an arbitrary vertex operator do \textit{not} contribute to scattering amplitudes.

\paragraph{Closed bosonic strings.} The spectrum of the closed bosonic string is given by tensoring that of the open under the level matching condition. The same applies to vertex operators. Therefore, we do not go into any detail and only recall the massless vertex operator as the most useful one
\begin{align}
    V_G(p,z,\ov{z})= \varepsilon_{\mu \nu}\, \partial X^{\mu}(z) \, \ov{\partial} X^{\nu}(\ov{z}) \, e^{ip \cdot X(z,\ov{z})}\,,
\end{align}
where decomposing $\varepsilon_{\mu \nu}$ into $so(d-2)$ irreps yields the graviton, Kalb--Ramond and dilaton degrees of freedom. 

\subsection{String spectrum's Zoology}
\label{subsec:zoology}

Let us now consider the string spectrum from different angles.  In principle, one can work out the spectrum's decomposition into irreducible representations of Wigner's little algebra \textit{level--by--level}, namely all states at a given $N$ are enumerated and built covariantly, as reviewed in the previous Subsection, or in the light--cone gauge, see Appendix \ref{app:lightcorn}. While it is widely appreciated that this procedure becomes technically cumbersome for very high $N$, the characters of various string spectra have been explicitly constructed, which reduces the problem to that of expressing a given character as a sum of characters of irreducible representations, which can be solved level--by--level, if needed \cite{Hanany:2010da}. Let us also highlight that it is possible to show that the Hilbert space induced by the BRST cohomology is free from negative--norm states, without having to enumerate all physical states explicitly \cite{Kato:1982im,Hwang:1982mc,Spiegelglas:1986xe,Henneaux:1986kp,Freeman:1986fx}. 

A generic on--shell (massive) physical state is associated with a polarization tensor $$\epsilon^{\mu_1(s_1),\, \mu_2 (s_2),...,\, \mu_k(s_n)}\,(p)\,,$$ 
where we use commas to separate groups of symmetric indices. It is an irreducible Lorentz tensor, namely traceless with a definite type of Young symmetry; it is also $p$--transverse in all Lorentz indices. Any physical string state can thus be represented by a Young diagram whose lengths of rows in numbers of boxes are given by the labels $s_1,\ldots,s_n$ as\footnote{we will come back to the irreducibility conditions of tensors and Young diagrams in section \ref{sec:interlude}}
\begin{align} \label{generalYD}
   \YY{s_1,\ldots, s_n} :\quad  \RectDRow{6}{5}{4}{2}{$s_1$}{$s_2$}{$\cdots$}{$s_n$}
    \quad , \quad s_1 \geq  s_2 \geq \ldots \geq s_n\,.
\end{align}
Any such diagram is part of the physical string spectrum and there are \textit{infinitely} many ways to embed it into the spectrum, in other words to dress a given polarization tensor with a polynomial $F$ such that it represents the vertex operator of a physical state: the string spectrum is degenerate in spin.
The open bosonic string spectrum up to $N=6$ \cite{Manes:1988gz, Hanany:2010da} is displayed in table \ref{firstseven}. Here the only peculiarity is that the vector at $N=1$ is massless, with all other states being massive. In red what is traditionally referred to as the leading Regge trajectory, namely the set of highest--spin states per level, is highlighted. The question then of how to order the subleading ones becomes one of ordering Young diagrams, since they encode the weights of $so(d-1)$, namely ultimately spin. A possible ordering is by decreasing length of rows starting from the first, as in table \ref{firstseven}; for example, the set of states highlighted in blue can be referred to as the first subleading trajectory. Another way of grouping states together could be by the maximum weight $K$ of the descendants of $\pl X$ they necessitate to be constructed, see Appendix \ref{app:lightcorn} for more detail in the oscillator language. For example, $K=1$ for the leading Regge trajectory.  

\begin{table}
\centering 
\renewcommand{\arraystretch}{1.5}
  \begin{tabular}{ c || l  }
   $N$ & decomposition in physical states  \\ \hline \hline
   $0$ & $\textcolor{red}{\bullet}$ \\
   $1$ & $\textcolor{red}{\YoungpA_{so(d-2)}}$ \\
   $2$ & $\textcolor{red}{\YoungpB}$ \\
   $3$ & $\textcolor{red}{\YoungpC} \oplus \textcolor{blue}{\YoungpAA}$ \\
   $4$ & $\textcolor{red}{\YoungpD} \oplus \textcolor{blue}{\YoungpBA} \oplus \textcolor{olive}{\YoungpB} \oplus \bullet$ \\
   $5$ & $\textcolor{red}{\YoungpE} \oplus \textcolor{blue}{\YoungpCA} \oplus \textcolor{olive}{\YoungpC} \oplus \YoungpBA \oplus \YoungpAA\oplus \YoungpA$  \\
   $6$ & $\textcolor{red}{\YoungpF} \oplus \textcolor{blue}{\YoungpDA} \oplus \textcolor{olive}{\YoungpD} \oplus \YoungpCA\oplus \YoungC  \oplus \textcolor{violet}{\YoungpBB} \oplus \YoungpBA  \oplus 2\,\YoungpB \oplus \textcolor{teal}{\YoungpAAA} \oplus \YoungpA\oplus \bullet$
  \end{tabular}
\renewcommand{\arraystretch}{1}
\caption{Open bosonic string, physical content per level up to $N=6$.} \label{firstseven}
\end{table}

In this work, instead of proceeding level--by--level or by Regge trajectories, we approach the construction of the spectrum in an alternative manner. Let us first define the notion of depth, anticipating its applications in section \ref{sec:excavation}. It is easy to see that the lowest level $N_{\text{min}}$ at which a physical state $\YY{s_1,\ldots, s_n}$ can (and will) occur is 
\begin{align} \label{minN}
    N_{\text{min}}=\sum_{i=1}^n s_i \,i\,.
\end{align}
It may also be possible to find such a state at level $N=N_{\text{min}}+w$, so let us define
\begin{align} \label{depthdef}
    w \equiv N-\sum_{i=1}^n s_i i
\end{align}
and refer to $w$ as depth in what follows. We can now consider families of trajectories \textit{depth--by--depth} and we list the first few families in table \ref{depth_traj}, with the number displayed in the first row of a Young diagram being the value of $s_1$, i.e. the spin in case of totally symmetric tensors, in such a way that $N=s$ for any state. For example, the family $w=0$ consists of the leading and subleading Regge trajectories, as well as (infinitely many) subsubleading trajectories, the lightest states of which are highlighted in violet and teal in table \ref{firstseven}. Another example is the subsubleading Regge trajectory highlighted in olive in table \ref{firstseven}, which becomes the first member of the $w=2$ family in table \ref{depth_traj}. It should be noted that, in table \ref{depth_traj}, not all trajectories listed start at the lowest possible level where the displayed Young diagram makes sense. For instance, the subsubleading trajectory $\RectARow{5}{$s-2$}$ does not start at level $s=2$, but at $s=4$. Likewise, $\RectARow{4}{$s-4$}$ has multiplicity $2$ at a sufficiently high level, but the two components start at different levels ($4$ and $6$). 
\begin{table}
\centering 
\renewcommand{\arraystretch}{1.5}
  \begin{tabular}{ c || l  }
   $w$ &  trajectories \\ \hline \hline
   $0$ & $\textcolor{red}{\RectARow{6}{$s$}} \oplus \textcolor{blue}{\RectBRow{5}{1}{$s-2$}{}} \oplus \textcolor{violet}{\RectARowBlock{4}{$s-4$}{\BlockA{2}{1}}} \oplus \textcolor{teal}{\RectARowBlock{4}{$s-5$}{\BlockA{1}{2}}}\oplus\RectARowBlock{4}{$s-6$}{\BlockA{3}{1}}\oplus \ldots$ \\
   $1$ & $\RectBRow{4}{1}{$s-3$}{}\oplus \RectARowBlock{4}{$s-5$}{\BlockA{2}{1}} \oplus \ldots$ \\
   $2$ & $\textcolor{olive}{\RectARow{5}{$s-2$}} \oplus 2^*\,\RectBRow{4}{1}{$s-4$}{}\oplus \ldots$ \\
   $3$ & $\RectARow{4}{$s-3$}\oplus \ldots$ \\
   $4$ & $3^*\,\RectARow{4}{$s-4$} \oplus \ldots$ \\
  \end{tabular}
\renewcommand{\arraystretch}{1}
\caption{Open bosonic string, physical families per depth up to $w=4$.} \label{depth_traj}
\end{table}

The spectrum displays certain patterns; trajectories, once having emerged, do not disappear: once a certain Young shape appears, it is the first state of a new trajectory, proceeding along which is about increasing the length of the first row. In fact, it turns out that for any given state one can find states at higher levels whose polarization tensors are obtained by adding boxes to any of the rows. The first row is only distinguished by the fact that one can immediately add any number of boxes and that adding one box increases the level by one. Adding a box to the second row can be done as long as its length does not exceed the first row and each box from the second row costs two units of energy, etc. This motivates us to introduce the notion of the family of trajectories that groups trajectories with polarization tensors having Young diagrams with $n$ rows.

At this point it appears that, to fully determine a physical state, three pieces of information are required: its polarization tensor, its level $N$ \textit{or} depth $w$ and its multiplicity. The full spectrum looks complicated and its organizing principles are not manifest: the main difficulty is to determine what kind of and where trajectories emerge, namely which Young diagrams appear at what level, while following them along presents no difficulty. In what follows, we will use the maximum weight $K$ of the descendants of $\pl X$ in physical polynomials to organize and probe trajectories \textit{depth--by--depth} in a covariant way that will prove to be advantageous in efficiency compared to the level--by--level approach, as well as show the existence of a certain subset of trajectories. Let us note that $w$ naturally groups trajectories within families, to which families we will also be referring with the term ``trajectories'' and that $w$ can also be thought of as a measure of the complexity of a state, but for now it is just a way of organizing trajectories.

\section{Excavating string trajectories}
\label{sec:excavation}

In this section, we probe physical vertex operators by $K$, namely the maximal weight of the descendants of $\pl X$ their polynomials $F$ involve; the value of $K$ appears as an index of $F$. We use the symbol $\mathbb{V}$ to denote such operators that describe whole trajectories, while we keep the symbol $V$ for whole levels or individual states. 

\subsection{Warm--up: the leading Regge trajectory}
\label{sec:leading}

Let us begin with the simplest case of the polynomial $F$ of a generic vertex operator (\ref{generic_vo_polynomial}), namely the one in which $F$ depends only on the first derivative of $X\,$, i.e. $K=1$:
\begin{align} \label{ansatz_leading}
   \mathbb V_{F_1}(z)&= F_1(\pl X)\, e^{i p\cdot X}= \sum_s F_{\mu_1...\mu_s}(p) \, \pl X^{\mu_1}(z) \ldots \pl X^{\mu_s}(z)\, e^{i p\cdot X(z)}\,,
\end{align}
where $s \in \mathbb{N}$ is a priori free and the rank--$s$ tensors $F^{\mu_1...\mu_s}$ are by construction totally symmetric. Then (\ref{ope_general_physical}) is satisfied identically at the level of the pole of order $1$, while the poles of order $2$, $3$ and $4$ enforce respectively, see e.g. \cite{Sagnotti:2010at},
\begin{align} \label{first_regge_constraints}
\big( \alpha'p^2-1 +N_1 \big) F_1=0 \quad , \quad p \cdot \frac{\delta F_1}{\delta \partial X}=0 \quad \textrm{and} \quad \frac{\delta}{\delta \partial X  }\cdot\frac{\delta}{\delta \partial X  }F_1 =0\,,
\end{align}
where the operator $N_1$ is defined as
\begin{align}
    N_1 \equiv \pl X \cdot  \frac{\delta}{\delta \pl X}  \quad \Rightarrow \quad N_1 F_1 = s\, F_1\,,
\end{align}
which matches the definition (\ref{numberop}) following Appendix \ref{app:alphas} and in the last formula $F$ was restricted to spin--$s$.

The first of (\ref{first_regge_constraints}) fixes the mass, i.e. $p^2$,  of the state in terms of the number of $\pl X$ in $F$, which here is directly the spin. Therefore, strictly speaking, one should write the expansion (\ref{ansatz_leading}) as
\begin{align} \label{ansatz_leading_strict}
    \mathbb  V_{F_1}(z)&=  \sum_s F_{\mu_1...\mu_s}(p_s) \, \pl X^{\mu_1}(z)\ldots \pl X^{\mu_s}(z) \, \exp{i p_s\cdot X(z)}\,,
\end{align}
stressing that different spins have $p^\mu_s$ belonging to different mass--shells; we will avoid such an unpleasant notation and always use just one $p^\mu$ that will land on the right mass--shell whenever needed. $\mathbb V_{F_1}$ along with the on--shell conditions (\ref{first_regge_constraints}) can thus be thought of as the vertex operator creating the entire leading Regge trajectory, highlighted in red in tables \ref{firstseven} and \ref{depth_traj}. The spin-one state is massless and the BRST--exact terms manifest the gauge symmetry, (\ref{vector_inv_comm}). All other states are massive and physical, i.e. there are no BRST--exact terms to mod out. The last two conditions of (\ref{first_regge_constraints}) imply that the Taylor coefficients $F^{\mu_1...\mu_s}$ are $p$--transverse and traceless. As it is customary in the literature, one can solve these with the help of an auxiliary vector $\epsilon^\mu$ such that $\epsilon \cdot p=0$ and $\epsilon\cdot \epsilon=0$. In this way, for any function $f(x)\,$,
\begin{align}
   \mathbb  V_{F_1}&= f(\epsilon\cdot \pl X)\, e^{i p\cdot X}
\end{align}
is a physical vertex operator. Bearing in mind that Wick's theorem favors exponentials, it makes sense to choose $f(x)=e^{ix}$ to rewrite \cite{Kawai:1985xq}
\begin{align} \label{gen_xi}
  \mathbb   V_{F_1}(z,p,\epsilon)&= \exp\big(i p\cdot X + i \epsilon \cdot \pl X \big) \,.
\end{align}

\subsection{All trajectories at once}
The most general vertex operator can be written as
\begin{align} \label{generic_form}
   \mathbb V_F(z,p)&= F\big[\pl X(z), \pl^2 X(z),....\big] e^{i p\cdot X(z)}\,,
\end{align}
where the generating function $F$ is responsible for making it BRST--closed and primary of weight $1$. Now $K$ is a free index, so we suppress it: the form (\ref{generic_form}) has the potential to probe \textit{all} trajectories of the open bosonic string. At this point it is convenient to introduce the shorthand notation
\begin{align}
    X^{(n)}_\mu(z) \equiv \pl^n X_\mu(z)\,.
\end{align}
Taking the OPE of any operator $ \mathbb V$ with $T$ can be implemented via 
\begin{align} \label{general_generating}
    :T(z):\,:  \mathbb V(w): &=\exp{ -2\alpha'\,\eta_{\mu\nu} \sum_{n=0} \frac{\delta}{\delta X^{(1)}_\mu(z)} \frac{n!}{(z-w)^{n+1}} \frac{\delta }{\delta X^{(n)}_\nu(w)} }\, :T(z)\,  \mathbb V(w):\,,
\end{align}
where $X^{(k)}$ are treated as independent variables. We now suppress the normal ordering symbol as there is no confusion. The OPE of any operator with the energy--momentum tensor (\ref{EM}) truncates at the second term after expanding the exponent due to $T$ being quadratic in $X^{(1)}$. It is also convenient to split the sum into $n=0$ and $n>0$ parts since $X\equiv X^{(0)}$ contributes in a way different from $X^{(n>0)}$. Using (\ref{general_generating}), we find that
\begin{align} \label{calcul_ope}
\begin{aligned}
    T(z)\,\mathbb V_F(w) \, &\sim \bigg\{ \frac{1}{z-w} i \, p \cdot X^{(1)}(w) \,F(w) + \frac{1}{(z-w)^2} \,  \alpha' p^2 \,F(w)
\crbig
& \quad  +\displaystyle \sum_{n=1} \frac{n!}{(z-w)^{n+1}} \bigg[  X^{(1)}(w) \cdot \frac{\delta F}{\delta X^{(n)}(w)}-2\alpha' \frac{1}{z-w} \, i \, p \cdot \frac{\delta F}{\delta X^{(n)}(w)}
\crbig
& \quad  \quad +\displaystyle \sum_{m=1}  \bigg( \frac{(z-w)^m}{m!} X^{(m+1)}(w) \cdot \frac{\delta F}{\delta X^{(n)}(w)}
  \crbig
& \quad \quad\quad -\alpha' \frac{m!}{(z-w)^{m+1}} \frac{\delta^2 F}{\delta X^{(n)}(w) \cdot \delta X^{(m)}(w)} \bigg) \, \bigg] \,\bigg\} e^{ip\cdot X^{(0)}(w)}\,.
\end{aligned}
\end{align}
Comparing this result (\ref{calcul_ope}) with the general form of the OPE (\ref{ope_general_physical}), namely\footnote{The second bracket unfolds $\pl \mathbb V_{F}$.}
\begin{align}
    T(z)\,\mathbb V_F(w) \, \sim \frac{\mathbb V_F(w)}{(z-w)^2}+\frac{1}{z-w}\bigg[ip\cdot X^{(1)}(w) \mathbb V_F(w)+ \sum_{n=1} X^{(n+1)} \cdot \frac{\delta F}{\delta X^{(n)}(w)}\, e^{ip\cdot X^{(0)}(w)}\bigg]
\end{align}
and imposing that they be equal for poles of every order, we find that equality at the level of the pole of order $1$ is trivially satisfied,  while those of order $2$ and $n+2>2$ yield respectively
\begin{align} \label{mass_covariant}
    (L_0-1)F=\bigg(\sum_{n=0} n \, X^{(n)}\cdot \frac{\delta}{\delta X^{(n)}} +\alpha' p^2 -1\bigg) \,F&=0\,,
\end{align}
where 
\begin{align} \label{numbernew}
    T\fud{n}{n}\equiv N_n=X^{(n)}\cdot  \frac{\delta}{\delta X^{(n)}}
\end{align}
is the number operator in accordance with the definition (\ref{numberop}), cf. Appendix \ref{app:alphas}, and
\begin{align}\label{general_eq_covariant}
\begin{aligned}
L_nF&=\bigg[ 2\alpha' \, n! \,ip \cdot \frac{\delta }{\delta X^{(n)}} + \alpha' \displaystyle \sum_{m=1}^{m=n-1} m! (n-m)! \frac{\delta^2 }{\delta X^{(m)}\cdot \delta X^{(n-m)}} 
\crbig
& \qquad\qquad - \displaystyle \sum_{m=0} \frac{(n+m+1)!}{m!} X^{(m+1)} \cdot \frac{\delta }{\delta X^{(n+m+1)}} \bigg] F =0 \quad , \quad \forall \, n \in \mathbb{N}^* \,. 
\end{aligned}
\end{align}
(\ref{mass_covariant}) is the on--shell mass constraint. It is straightforward to see that (\ref{mass_covariant}) and (\ref{general_eq_covariant}) reproduce the constraints (\ref{first_regge_constraints}) for the first Regge trajectory, in particular the last two after setting $n=1$ and $n=2$ respectively in the second of (\ref{general_eq_covariant}); for $n>3$, the latter yields the triviality $0=0$ for the trajectory in question.

The above calculation is, of course, homotopic to directly imposing the physical state condition (\ref{GBphys}) à la Gupta--Bleuler: the differential operators that act on $F$ in (\ref{mass_covariant}) and (\ref{general_eq_covariant}) represent the $L_n$'s in the operator formalism in a ``1--1'' fashion, as we show in Appendix \ref{app:alphas}. In both languages, the Virasoro constraints suggest defining the following atomic structures
\begin{align}\label{atomst}
     T\fud{k}{l}\equiv X^{(k)} \cdot\frac{\delta }{\delta X^{(l)}} \quad , \quad T_{kl}&\equiv \frac{\delta^2 }{\delta X^{(l)}\cdot \delta X^{(l)}}\,, 
\end{align}
which we may also write as
\begin{align}
    T\fud{k}{l}= X^{(k)} \cdot  P_{(l)} \quad , \quad T_{kl}&= \eta_{\mu\nu} P_{(k)}^\mu P_{(l)}^{\vphantom{\mu}\nu}  \quad , \quad  P_{(l)}^\mu\equiv \frac{\pl}{\pl X^{(l)}_\mu}\,.
\end{align}
Together with a few other operators, they form an $sp(2K)$--algebra, where $K$ has become the maximal value of the labels $k,l,m,n$. This algebra commutes with the Lorentz algebra and is instrumental in solving the Virasoro constraints, as we explain in the next Subsection. 
  
For completeness, let us note that for closed strings the most general generating function reads
\begin{align}
   \mathbb V_F(z,\ov{z},p)&= F\big[\pl X(z), \pl^2 X(z),\ldots,\ov{\pl} X(\ov{z}), \ov{\pl}^2 X(\ov{z}),\ldots\big] e^{i p\cdot X(z,\ov{z})}
\end{align}
and the Virasoro constraints are straightforward generalizations of those above.

\subsection{Interlude on irreducible tensors}
\label{sec:interlude}
The prefactor $F$ of a generic vertex operator $\mathbb V_F$ as in \eqref{generic_form} can be Taylor--expanded to reveal components of the form
\begin{align}
    F(X^{(1)},...)\ni F^{\mu(n_1)|....|\nu(n_k)}\, X^{(1)}_{\mu_1}...X^{(1)}_{\mu_{n_1}}\, ...X^{(k)}_{\nu_1}....X^{(k)}_{\nu_{n_k}}\,,
\end{align}
where we use ``$|$'' to separate groups of symmetric indices (with no additional algebraic constraints imposed a priori), so that a symmetric group $\mu_1\ldots \mu_k$ is abbreviated to $\mu(k)$. The Taylor coefficients $F^{\mu(n_1)|....|\nu(n_k)}$ are by construction very far from irreducible $so(d)$ tensors. They are also not $gl(d)$-irreducible. The latter must obey Young symmetry conditions, which ensure irreducibility under $gl(d)$, and the former be, in addition, $so(d)$--traceless (for $d$ even one can also impose self--duality constraints). Let us forget for a moment about vertex operators and briefly review these concepts while also introducing some useful notation.

\paragraph{$\boldsymbol{gl(d)}$ tensors.} It is convenient to define operators that impose Young symmetry conditions and tracelessness. For a generic function $F=F(X^{(1)},\ldots)$ to represent an irreducible $gl(d)$ tensor, the Young symmetry condition has to be imposed, which we may write as
\begin{align} \label{YS}
    T\fud{k}{l} F&=0 \,, \qquad k<l\,,
\end{align}
using the operators defined in (\ref{atomst}). They imply that symmetrization of all indices from the $k$--group with one index from any of the following groups, $l>k$, must give zero. The conditions (\ref{YS}) are not independent from each other and one can choose a smaller subset to impose that still makes the tensor $F$ irreducible under $gl(d)$, which corresponds to the simple roots of the Howe dual algebra, as we will see below. For $K=1$, there are no Young conditions and all tensors within $F_1$ are just symmetric. The first nontrivial case is $K=2$, in which $F_2$ can be expanded as
\begin{align}
    F_2(X^{(1)},X^{(2)})=\sum_{k, l} F^{\mu(k)|\nu(l)}\, X^{(1)}_{\mu_1}\ldots  X^{(1)}_{\mu_k}\, X^{(2)}_{\nu_1}\ldots  X^{(2)}_{\nu_{l}}\,.
\end{align}
Irreducible tensors under $gl(d)$ are then singled out by
\begin{align}
  T\fud{1}{2} F_2=0 \quad \Rightarrow \quad   F^{\mu(k),\mu \nu(l-1)}&=0 \,,
\end{align}
where we use commas to separate groups of symmetric indices that also obey the Young conditions and all indices that are to be symmetrized are denoted by the same letter, i.e. $\mu(k), \mu \equiv (\mu_1 , \ldots,\mu_k,\mu_{k+1})$. Every tensor can be decomposed into irreducible ones. However, each type of an irreducible tensor may lead to more than one polynomial. For example, with the same irreducible Taylor coefficients we can write
\begin{align}\label{genericirr}
    F_2(X^{(1)},X^{(2)})=\sum_{\substack{k\geq l\\ k\geq l+i }} F^{\mu(k-i)\nu(i),\nu(l)}_i\, X^{(1)}_{\mu_1}\ldots  X^{(1)}_{\mu_{k-i}}\, X^{(2)}_{\nu_1}\ldots  X^{(2)}_{\nu_{l+i}}\,.
\end{align}

Interestingly, the operators $T\fud{k}{l}$ form the algebra $gl(\bullet)$, where $\bullet$ is the range of the indices $k,l,...$. In particular, using the definitions (\ref{atomst}) we obtain
\begin{align} \label{algebra_1}
    [T\fud{k}{l},T\fud{m}{n}]&= \delta_l^m T\fud{k}{n}-\delta^k_n T\fud{m}{l}\,.
\end{align}
This algebra commutes with $gl(d)$ that we can realize on indices $\mu,\nu,...$
\begin{align}
[gl(d),gl(\bullet)]=0    
\end{align}
and they are the maximal algebras with this property: these two algebras are called \textit{Howe duals} \cite{Howe}, see \cite{Basile:2020gqi} for a review. The Cartan generators are $H_k\equiv T\fud{k}{k}$ (no summation) and count the number of $X^{(k)}$ present in a generating function. One can choose
\begin{align}
    a^I&= \{ a^{k,l}=T\fud{k}{l} \quad ,\quad k<l \}\,, &
    a_I^\dag&= \{a^\dag_{k,l}=T\fud{k}{l} \quad , \quad k>l\}
\end{align}
be respectively lowering and raising generators of $gl(\bullet)$. It is obvious that any polynomial $f$  can be decomposed into eigenvectors of $T\fud{k}{k}$, i.e. those having a fixed number of $X^{(k)}$. Any such eigenvector can further be decomposed into 
\begin{align}\label{creatingpoly}
    f= \sum a^\dag_{I_1}...a^\dag_{I_m} f_Y \quad , \quad  a^{I} f_Y=0\,,
\end{align}
c.f. \eqref{genericirr}. Howe duality leads to a very useful statement that the lowest weight conditions
\begin{align} \label{lws}
    a^{I} f_Y=0 \quad , \quad H_i f_Y= s_i f_Y
\end{align} with respect to $gl(\bullet)$ imply that $f_Y$ is an irreducible $gl(d)$ tensor:
\begin{align}
    &\text{l.w.s. of }gl(\bullet) \quad \Longleftrightarrow \quad \text{irreps of } gl(d)\,.
\end{align}
If the eigenvalues of the Cartan generators are fixed, $f_Y$ is a single irreducible representation of $gl(d)$, whose Young diagram is $\YY{s_1,\ldots s_n}$. Otherwise, $f_Y$, when expanded, covers all  finite dimensional irreps of $gl(d)$ with Young diagrams having at most $K$ rows, $\bullet=K$. 

Coming back to the example with $K=2$, the lowest weight vectors correspond to tensors with (no more than) two groups of indices, each coefficient obeying the Young conditions
\begin{align}
     a^1 F_2\equiv T\fud{1}{2}F_2=0\,.
\end{align}
The creation operator is $a^\dag_1 \equiv T\fud{2}{1}$, which allows to reach other arrangements of $X^{(1)}$, $X^{(2)}$ on the same tensors, c.f. \eqref{genericirr}
\begin{align}
   \sum \tfrac{k!}{(k-i)!}F^{\mu(k-i)\nu(i),\nu(l)}_i\, X^{(1)}_{\mu_1}\ldots  X^{(1)}_{\mu_{k-i}}\, X^{(2)}_{\nu_1}\ldots  X^{(2)}_{\nu_{l+i}}&= (a^\dag)^i F\,.
\end{align}
We can split the Howe dual $gl(2)$ into $gl(1)$, generated by $T\fud{1}{1}+T\fud{2}{2}$, and $sl(2)$, formed by $a^1=T\fud{1}{2}$, $a^\dag_1=T\fud{2}{1}$ and $h=T\fud{1}{1}-T\fud{2}{2}$. In this way, totally symmetric tensors ($l=0$, $\YY{k}$) correspond to the $(k+1)$--dimensional representations of $sl(2)$ and type--$\YY{k,l}$ tensors form the $\big((k-l)+1\big)$--dimensional representation of $sl(2)$. Indeed, 
\begin{align}
    a^1F_{k,l}=0 \quad , \quad (a^\dag_1)^{k-l+1} F_{k,l}=0\,.
\end{align}
The singlets of $sl(2)$ correspond to rectangular Young diagrams, $k=l$. Extending the 2--row example to the most general case is straightforward. 

\paragraph{$\boldsymbol{so(d)}$ tensors.} Irreducible $gl(d)$ tensors are a prelude to $so(d)$ ones. In order for $F$ to represent an irreducible $so(d)$ tensor, one has to further impose a tracelessness constraint via
\begin{align} \label{traceless}
    T_{kl} F &=0 \,.
\end{align} 
Again, not all of these conditions are independent. In fact, one can generate them all from $T_{11}F=0$ by means of the Young symmetry condition and the algebra
\begin{align}
        [T_{km}, T\fud{l}{n}] = \delta^l_k T_{mn}+\delta^l_m T_{kn}\,,
\end{align}
which can be derived using the definition (\ref{atomst}). If the dimension $d$ is even and $f$ represents a tensor with a symmetry of a Young diagram with $d/2$--rows, one can further impose (anti)--self duality constraints with the help of the Levi--Civita symbol; we ignore this possibility since there is no parity violation in the spectrum of bosonic strings.  

We may also extend the set of operators (\ref{atomst}) with trace--creation ones
\begin{align}
    T^{km}=X^{(k)}\cdot X^{(m)} \,.
\end{align}
The complete set of nontrivial commutation relations involving the generators $T\fud{k}{l}$, $T^{kl}$, $T_{kl}$ reads
\begin{align}
    [T\fud{l}{n},T^{km}]&= \delta^k_n T^{lm}+\delta^m_n T^{lk}\\
    [T_{km}, T\fud{l}{n}]&= \delta^l_k T_{mn}+\delta^l_m T_{kn}\\
    [T\fud{k}{l},T\fud{m}{n}]&= \delta_l^m T\fud{k}{n}-\delta^k_n T\fud{m}{l}\\
    [T_{km}, T^{ln}]&= d(\delta_k^n \delta_m^l + \delta_k^l \delta_m^n) + \delta^l_k T\fud{n}{m}+\delta^l_m T\fud{n}{k}+\delta^n_k T\fud{l}{m}+\delta^n_m T\fud{l}{k}\label{traceantitrace}\,,
\end{align}
where in the last line the ``central term'' can be absorbed in the definition of $T\fud{k}{l}$ via  $T\fud{k}{l}\rightarrow T\fud{k}{l}+\delta^k_l d/2$. The resulting algebra is $sp(2\bullet)$. This algebra commutes with the orthogonal algebra $so(d)$ realized on indices $\mu, \nu, \ldots$. In practice, the real form we are interested in is the Lorentz algebra $so(d-1,1)$. The creation operators get extended by $T^{km}$ and the annihilation operators by $T_{km}$, namely now
\begin{align}
    a^\dag_I=\{T^{km}; T\fud{k}{l}, k>l\} \quad , \quad a^I=\{T_{km}; T\fud{k}{l}, k<l\}\,.
\end{align}
The implication of Howe duality is that
\begin{align}
    &\text{l.w.s. of }sp(2\bullet) \quad \Longleftrightarrow \quad \text{irreps of } so(d-1,1)\,,
\end{align}
where l.w.s. satisfy \eqref{lws}. Now, the most general polynomial can be reduced to irreducible $so(d-1,1)$ tensors, which can then be dressed with creation operators as in \eqref{creatingpoly}. The latter include traces, e.g. schematically
\begin{align}
    F_2(X^{(1)},X^{(2)})=\sum_{} \eta^{\mu\mu}...\eta^{\mu\nu}...\eta^{\nu\nu}...F^{\mu(k-i)\nu(i),\nu(l)}\, X^{(1)}_{\mu_1}\ldots  X^{(1)}_{\mu_{.}}\, X^{(2)}_{\mu_1}\ldots  X^{(2)}_{\mu_{.}}\,.
\end{align}

\paragraph{Polarization tensors of string theory.} It is worth stressing that the discussion of irreducibility constraints above does not yet concern string theory per se. In particular, even though they are represented by $p$--transverse and $so(d-1,1)$ irreducible polarization tensors, most of the physical states of bosonic string theory will have complicated representations as polynomials in $X^{(k)}_\mu$. This is exactly the problem we address in this work:  given a polarization tensor represented by a function $F$, how can we render it physical by embedding it at a specific depth $w$ in the spectrum? 

\begin{table}
\centering 
\renewcommand{\arraystretch}{1.5}
  \begin{tabular}{ c || c | c }
   generators & algebra & Howe dual algebra \\ \hline \hline
   $T\fud{k}{l}$ & $gl(\bullet)$ & $gl(d)$ \\ \hline
   $T\fud{k}{l}$, $T_{kl}$, $T^{kl}$ & $sp(2\bullet)$ & $so(d-1,1)$ \\ \hline
   \multirow{2}{*}{$T\fud{0}{l}, T\fud{k}{l}$, $T_{kl}$, $T^{kl}$} & parabolic subalgebra of & \multirow{2}{*}{$iso(d-1,1)$} \\
   & $sp\big(2(\bullet+1)\big)$ &  
  \end{tabular}
\renewcommand{\arraystretch}{1}
\caption{Algebras and their Howe duals ($k,l=1,2,\ldots$).} \label{table:algebrasH}
\end{table}
Let us first discuss how transversality can be imposed. Massive physical states are associated with $so(d-1)$ irreducible polarization tensors. Without breaking Lorentz covariance, one can single out such tensors as those irreducible under $so(d-1,1)$ and obeying the transversality constraint
\begin{align}\label{transverse}
    T\fud{0}{l} F=0 \quad , \quad l>0  \,,
\end{align}
where
\begin{align} \label{operator_3}
    T\fud{0}{l}\equiv p \cdot P_{(l)}= p \cdot \frac{\pl}{\pl X^{(l)}}\,.
\end{align}
The notation suggests that the momentum $p^\mu$ can be grouped with $X^{(k)}_\mu$ and, in fact, should be placed as the \textit{first element}. $T\fud{0}{l}$ extends the $sp(2\bullet)$ algebra of $T^{km}$, $T\fud{k}{l}$ and $T_{km}$ and together they form a parabolic subalgebra of $sp(2(\bullet+1))$ that commutes with the target space Poincar\'e algebra $iso(d-1,1)$. In table \ref{table:algebrasH} we summarize the above considerations on Howe duals. Let us notice that, if we add \textit{both} $p$ and $\pl/\pl p$ to the set of $X$ and $P$, we find an $sp(2(\bullet+1))$ algebra that does \textit{not} commute with $iso(d-1,1)$. 

In case of string theory, $\bullet=\infty$. The Virasoro algebra is realized as a subalgebra of $sp(2(\bullet+1))$, with respect to which the physical state conditions are the lowest--weight conditions. Had the Virasoro lowest--weight conditions corresponded to all annihilation operators of $sp(2\bullet)$, we would have ended up with all $so(d-1,1)$ irreducible tensors in terms of $X^{(k)}_\mu$. The difference between all polynomials in $X^{(k)}_\mu$ and those that are annihilated by $a^I$ of $sp(2\bullet)$ is that the latter must have $X^{(k)}_\mu$ consecutively filling the rows of Young diagrams, i.e. the indices corresponding to the first row are contracted with $X^{(1)}_\mu$'s, the indices of the second row with $X^{(2)}_\nu$, etc., e.g. for $\YY{s_1,\ldots, s_n}$ we can write
\begin{align}
    F_Y&= F^{\mu(s_1), \ldots, \nu(s_n)}\, X^{(1)}_{\mu_1}...X^{(1)}_{\mu_{s_1}}\, ...X^{(n)}_{\nu_1}....X^{(n)}_{\nu_{s_n}}\,. 
\end{align}
The $sp(2\bullet)$ l.w.s. condition is a much stronger condition than the Virasoro constraints. Nevertheless, it is easy to see that the $p$--transverse lowest--weight states of $sp(2\bullet)$ form an infinite subspace that solves the Virasoro constraints and contains complete trajectories. All other trajectories can be generated with the help of dressing functions or ``trajectory--shifting'' operators that are built out of the creation operators $a^\dag_I$ of the $sp(2\bullet)$ algebra, as we will demonstrate in explicit examples in the next Section.

\subsection{More on BRST cohomology}
With the integrated (\ref{generic_vo_polynomial}) and unintegrated (\ref{unintegrated}) vertex operators, the cohomology groups $H(Q)$ and $H(Q/d)$ can be associated accordingly; the generic vertex operators we treated using their integrated representatives (\ref{generic_form}). Using the operators defined in (\ref{numbernew}), (\ref{atomst}) and (\ref{operator_3}), the Virasoro constraints (\ref{mass_covariant}) and (\ref{general_eq_covariant}) we derived can be rewritten as
\begin{align} \label{constraints_new_form}
    (L_0-1)F&=\bigg(\sum_{n=0} n \, T\fud{n}{n} + \alpha'p^2 -1\bigg)F =0 \,,
    \crbig
    - L_nF&=\bigg[2i\alpha' \, n! \,T\fud{0}{n} + \alpha' \sum_{m=1}^{m=n-1} m! (n-m)! \,T_{m,n-m} -  T_n\bigg]F &=0\,, \qquad \forall \, n \in \mathbb{N}^* \,, \label{Lns}
\end{align}
where we have also defined
\begin{align}
    T_n&\equiv \sum_{m=0} \frac{(n+m+1)!}{m!} T\fud{m+1}{n+m+1} \quad , \quad n \in \mathbb{N}^*\,.
\end{align}
Using the algebra (\ref{algebra_1}), it is straightforward to find that
\begin{align}
    [T_n,T_k]= (n-k) \, T_{n+k} \quad , \quad    [L_n,L_k]= (n-k) \, L_{n+k}\,, \label{virasoro_new}
\end{align}
whose form is identical to the Virasoro algebra (\ref{vir_alg}) for positive integers; this is why we denote the differential operators that appear in the LHS of (\ref{Lns}) with the same symbol as the Virasoro operators.\footnote{Note that $L_n$ represented on operators have a particular structure: there is a first order differential operator, let's call it $D_1$, built of $T\fud{k}{l}$ with $l-k=n$ and the second order piece, let's call it $D_2$ built from $T_{m,n-m}$. It is clear that $[D_1,D_1]=D_1$ and $[D_1,D_2]=D_2$ with $D_1$ forming the same algebra as $L_n$ and $D_2$ being a module over it. Since the first order part $D_1$ of $L_n$ forms the same algebra as the complete $L_n$, one might be curious to see whether imposing $D_1$ leads to an interesting spectrum and what is the underlying theory, which is insensitive to the metric since $T\fud{k}{l}$ do not depend on the metric} As it was already mentioned in Section \ref{sec:ingredients}, while the number of constraints arising from (\ref{Lns}) seems to grow with $K$, only \textit{two} constraints, $L_1F=0$ and $L_2F=0\,$, are sufficient. 

As we will need the explicit form of the constraints (\ref{constraints_new_form}) and (\ref{Lns}) for $K=1,2,3$ in later sections, we gather the respective expressions below:
\begin{align}\label{leadingtrajex}
    \big( \alpha'p^2 -1+N_1 \big) \, F_1 =0 \quad , \quad     T^0_{\hphantom{0}1} F_1=0 \quad , \quad  T_{11}F_1=0 \,,
\end{align}
\begin{align} \label{VirasoroB}
    \big( \alpha'p^2 -1+T^1_{\hphantom{0}1} +2\, T^2_{\hphantom{0}2} \big) \,F_2 = 0 \quad , \quad
    \big( T^1_{\hphantom{1}2} - \alpha' i  \, T^0_{\hphantom{0}1} \big) F_2 =0  \quad , \quad
     \big(  T_{11}+4i \, T^0_{\hphantom{0}2} \big) F_2 =0,  
\end{align}
while $T_{12} F_2=0$, $T_{22}F_2=0$ follow from the commutators of $L_1$ and $L_2$, and
\begin{align}
    \big( \alpha'p^2 -1+T^1_{\hphantom{1}1} +2\,T^2_{\hphantom{2}2} + 3\, T^3_{\hphantom{3}3}  \big) \,F_3 &= 0 \label{simplest_eq_3_new_not_1}\,,
\crbig
\big( 3\, T^2_{\hphantom{2}3}  +  T^1_{\hphantom{1}2}   -\alpha' \, i \, T^0_{\hphantom{0}1} \big) \, F_3 &= 0\,, \label{simplest_eq_3_new_not_2}
\crbig
\big( \alpha'  T_{11} +4\alpha' \, i\, T^0_{\hphantom{0}2}  -6 T^1_{\hphantom{1}3} \big) \, F_3 &= 0\,, \label{simplest_eq_3_new_not_3}
\end{align}
which again imply a few other simpler conditions equivalent to $L_{3,4,5,6}$.

As explained in Section \ref{sec:ingredients}, the simplest BRST--exact terms can be written by means of any operator $U$ of weight $0$. Using generating functions, it should be chosen in the form
\begin{align} \label{arbitrary_op}
  \mathbb U(z,p)&= G\big[\pl X(z), \pl^2 X(z),....\big] e^{i p\cdot X(z)}\,.
\end{align}
Using (\ref{spurious_def}) and the OPE (\ref{general_generating}), this yields 
\begin{align} \label{spurious_new}
  \mathbb  V_{\textrm{sp}} (z,p) = D\, G\big[\pl X(z), \pl^2 X(z),....\big] \, e^{i p\cdot X(z)}
    \quad , \quad D  \equiv ip\cdot X^{(1)} + \sum_{n=1} T\fud{n+1}{n}\,,
\end{align}
where, to ensure that $\mathbb U$ has weight $0$ or equivalently $\mathbb V_{\textrm{sp}}$ weight $1$, the function $G$ must satisfy the same differential equation (\ref{general_eq_covariant}) as the function $F$ of physical states, but now the mass--level constraint becomes
\begin{align} \label{mass_covariant_spurious}
    \bigg(\sum_{n=0} n  \, T\fud{n}{n} +\alpha' p^2 \bigg) \,G&=0\,.
\end{align}
Comparing with (\ref{mass_covariant}),  we deduce that spurious states look like physical states at one level \textit{higher} than where they are supposed to be. We have already seen such an example: the (transverse) vector 
\begin{align} 
 \epsilon_\mu \, i\partial X^\mu(z) \, e^{ip\cdot X(z)} 
\end{align}
is physical at level $N=1$ and given by (\ref{open_vector}) but spurious at level $N=2$ as in (\ref{spurious_2_comm}). The action of $D$ corresponds to $L_{-1}$. Similarly, one can realize the action of $L_{-2}$, which together with $L_{-1}$ generate all spurious states. We will not be very concerned with BRST--exact terms, since there is a simple way to eliminate them by restricting to the transverse subspace.

\section{String spectrum's sections} 
\label{sec:slices}

In this Section, we work out how explicit examples of physical trajectories of different depth $w$, which were discussed in Subsection \ref{subsec:zoology}, can be embedded in the string spectrum using the covariant method developed in Section \ref{sec:excavation}. Since most of the open bosonic string spectrum is massive, we forget temporarily about the massless vector. With ``trajectory'' or ``family'' we refer to an infinite set of states that have similar symmetry types. For example, we will be considering symmetric states $\epsilon^{\mu(s)}$, i.e. spin--$s$, or $2$--row families, i.e. states $\epsilon^{\mu(s_1), \,\nu(s_2)}$ that enjoy the symmetries of $2$--row Young diagrams, and so on, as in the generic Young diagram \eqref{generalYD}.

\subsection{Embedding physical states} 

\paragraph{Principal embedding: $\boldsymbol{w=0}$.}  It is easy to see \cite{Weinberg:1985tv} that the simplest way to construct an operator that belongs to the string's BRST--cohomology is by contracting all indices from the $i$--th group of $\epsilon^{\mu(s_1), \ldots, \, \nu(s_K)}$ with $X^{(i)}_\mu$, namely by means of the polynomial
\begin{align} \label{pol_princ}
    F_\epsilon &= \epsilon^{\mu(s_1), \ldots, \, \nu(s_K)}\, X^{(1)}_{\mu_1}...X^{(1)}_{\mu_{s_1}}\, ...X^{(K)}_{\nu_1}....X^{(K)}_{\nu_{s_K}}\,.
\end{align}
(\ref{pol_princ}) may be referred to as the principal embedding, since it corresponds to the simplest vector that is the lowest weight state of Howe dual algebra $sp(2K)$. By construction, $F_\epsilon$ satisfies the Young symmetry condition (\ref{YS}) and is traceless (\ref{traceless}) and we also take it to be transverse (\ref{transverse}). It is clear then that $F_\epsilon$ is physical, since the operators $L_n$ in the Virasoro constraints \eqref{constraints_new_form} and (\ref{Lns}) are linear combinations of $T\fud{0}{k}$, $k>0$, that checks transversality,  and the annihilation operators $a^I$, which consist of $T\fud{k}{l}$ with $k<l$, that check Young symmetry, and $T_{m,n-m}$, that check tracelessness on various groups of indices. It is at this point that the notion of an \textit{infinite} family of states becomes natural: we can, in addition, fix explicitly the spin via $T\fud{i}{i}F_\epsilon=s_i F_\epsilon$ (no summation), but it is as easy to identify one such string state as to find infinitely many of them, since $s_i$ can be arbitrary. Given that string spectrum is repetitive, it is clear that the principal embedding gives states at the \textit{lowest} possible mass level where a given polarization tensor can occur. Consequently, the definition (\ref{depthdef}) of depth
yields
\begin{align}
      N=  N_{\text{min}}=\sum_{i=1}^K s_i \,i \quad , \quad w=0\,.
\end{align}
Evidently, all possible massive states, from scalars to arbitrarily complicated Young diagrams, appear at $w=0$.  

It is customary to factorize symmetric polarization tensors $\epsilon^{a(s)}$ as a product of polarization vectors $\epsilon^a$ 
\begin{align}
    \epsilon^{a(s)}=\epsilon^a\ldots \epsilon^a \quad , \quad \epsilon\cdot \epsilon=0\,,
\end{align}
with $\epsilon^a$ being light--like, which guarantees the tracelessness of $\epsilon^{a(s)}$. In order to build more complicated polarization tensors from elementary blocks, one can take a number of light--like vectors $\epsilon_a^{i}$, $i=1,\ldots, K$ that are orthogonal to each other; a polarization tensor with the symmetry of a Young diagram $\YY{s_1,\ldots, s_n}$ can then be written as
\begin{align}\label{YoungProjector}
    F^{a(s_1),\ldots, c(s_n)}&= \epsilon_1^{a_1}\wedge\ldots \wedge  \epsilon_n^{c_1}\, \epsilon_1^{a_2}\wedge\ldots \wedge  \epsilon_n^{c_2} \, \ldots \quad , \quad \epsilon^i\cdot\epsilon^j=0\,.
\end{align}
Here, $\wedge$ is the usual exterior product of vectors, whose purpose is to anti-symmetrize the tensor products. Once the polarization tensor is contracted with $X^{(k)}_a$, the symmetrization over the Lorentz indices is automatic. 

\paragraph{Non--principal embedding: $\boldsymbol{w>0}$.} 
To \textit{uplift} the principal embedding (\ref{pol_princ}) to a higher mass level $N>N_{\text{min}}$, or in other words \textit{increase its depth}, we have to dress it by a function $f$ constructed by means of the $sp(2(\bullet+1))$ creation operators $a^\dag_I$ as
\begin{align}
    F_\epsilon^f \equiv  f(a^\dag_I) \, F_\epsilon = f(T^{0m}, T^{mn}, T\fud{k}{l})\, F_\epsilon \quad , \quad k>l\,,
\end{align}
with the respective conformal weights related as 
\begin{align} \label{weightsdr}
    h_{ F_\epsilon^f} = h_f + h_{F_\epsilon} \quad \Rightarrow \quad h_f = N-N_{\text{min}}=w\,,
\end{align}
such that $F_\epsilon^f$ appears at level $N$. We may now think of the depth $w$ as the conformal weight of the uplifting operator $f$, namely the number of units of energy it adds to a trajectory at $N_{\text{min}}$. Let us note that the units of energy added by $T^{mn}$ and by $T\fud{k}{l}$ are $m+n$ and $k-l$ respectively. Since the polynomial $F_\epsilon^f $ of every trajectory employs a finite number $K$ of descendants of $X^{(1)}$, the Ansatz for $f$ is always finite and a single $f$ will give access to an entire trajectory. In other words, the dressing function $f$ shifts \textit{all} member--states of the trajectory $F_\epsilon$ of the principal embedding by $w$ units of energy up their respective $N_{\text{min}}$. Importantly, despite the fact that the dimension of the vector space spanned by $f(X^{(1)}_\mu, \ldots,X^{(K)}_\mu)$ grows with $N$, it always gives a (globally) bounded number of $so(d-1)$ irreducible representations, see also Appendix \ref{app:lightcorn}. In other words, for fixed $K$, one finds a family of a fixed number of trajectories (at sufficiently high $N$) and, in order to discover new trajectories, one has to increase $K$. 

In practice, we can write the most general Ansatz for $f$ that satisfies (\ref{weightsdr}) and can allow $F_\epsilon^f$ to represent a physical state and then impose $QF_\epsilon^f=0$, namely the Virasoro constraints (\ref{constraints_new_form}) and (\ref{Lns}). Similarly, we can write the most general Ansatz $F_\epsilon^g$ of conformal weight one and two units less than that of $F_\epsilon^f$, in order to determine the BRST--exact states $\delta F_\epsilon^f=  Q F_\epsilon^g$, $\delta f=Dg\sim L_{-1}g$ and those produced by $L_{-2}$, respectively. A useful observation is that, given a physical state $F^f_\epsilon$ represented by the Young diagram $\YY{s_1,\ldots, s_n}$, one can obtain states at higher levels by adding one box to any row, say the $i$--th, and increasing the level accordingly by $i$, while maintaining that the maximal length of every row is that of its preceding one. This is manifest in the structure of the dressing functions $f$, whose coefficients will depend on the lengths $s_j$ in a smooth way, as we will see later in specific examples. 

\paragraph{Restricting to the transverse subspace.} One simplifying assumption is that vertex operators can be restricted to depend on the transverse metric\footnote{One of the proofs of positivity of the physical states relies on mapping physical states into the transverse subspace, see e.g. \cite{Scherk:1974jj}. One can give a simpler argument that applies to all string theories. Indeed, $L_n$ begins with $T\fud{0}{n}=p\cdot \pl/\pl X^{(n)}$ and one can reduce the system to the cohomology of this set of commuting operators, which leads to the transverse Ansatz. More formally, $L_n=T\fud{0}{n}+l_n$. Operators $T\fud{0}{n}$ form an abelian subalgebra $\mathfrak{a}$ and $l_n$ form a subalgebra $\mathfrak{h}$, such that $[\mathfrak{h},\mathfrak{a}]\in \mathfrak{a}$, $[\mathfrak{h},\mathfrak{h}]\in \mathfrak{h}$. Therefore, one can use Hochschild-Serre spectral sequence that reduces first the problem to the cohomology of $\mathfrak{a}$. The latter are just transverse tensors. The induced differential of $\mathfrak{h}$ on this subspace is not just a restriction, but has certain correction terms to preserve the transverse subspace, the net result being that the trace creation/annihilation operators of $sp(2K)$ needs to be taken with respect to the transverse metric $\eta_\perp$. Even simpler, $p$ can be assigned degree $1$, which separates $T\fud{0}{n}$ from the rest and allows one to compute its cohomology first.} only, see \cite{Manes:1988gz} for bosonic strings. Within our formalism using dressing functions $f$, this means that, instead of $\eta_{\mu\nu}$, we can use
\begin{align}
    \eta_\perp^{\mu\nu}&= \eta^{\mu\nu} - \frac{p^\mu p^\nu}{p^2}\,.
\end{align}
Among the set of operators $\{T\fud{0}{l}, T\fud{k}{l}$, $T_{kl}$, $T^{kl}\}$, this affects \textit{only }the trace--creation operators $T^{kl}$, which can be replaced with
\begin{align} \label{tracecreatetr}
    T^{kl}_\perp & = \eta_\perp^{\mu\nu}X^{(k)}_\mu X^{(l)}_\nu\,,
\end{align}
which essentially reduces $sp(2(K+1))$ to $sp(2K)$. Consequently, the Virasoro constraint (\ref{constraints_new_form}), namely the mass--shell condition, remains unchanged, while (\ref{Lns}) simplifies to
\begin{align} \label{constraints_new_formB}
    - L_n^\perp F&=\bigg[ \alpha' \sum_{m=1}^{m=n-1} m! (n-m)! \,T_{m,n-m}^\perp -  T_n\bigg]F =0\,, \qquad \forall \, n \in \mathbb{N}^* \,.
\end{align}
In addition, one needs to replace $d$ with $d-1$ in the commutator \eqref{traceantitrace}. Crucially, the physical polynomials are now free from terms \textit{linear} in the momentum $p$, there are namely \textit{no} BRST--exact states in the transverse target space!

As a last comment, while the dressing function $f$ can depend on the spacetime dimension $d$, there is nothing special about the critical dimension $d=26$ at the moment.

\subsection{Principal embedding}

As it was discussed previously, there are infinitely many trajectories, each corresponding to a different value of $K$, that correspond to the principal embedding $w=0$, for which the dressing functions are trivial. We now treat the examples of such trajectories displayed in the $w=0$ row of table \ref{depth_traj}.

\paragraph{$\boldsymbol{K=1}$.} We begin with the simplest case of polynomials $F$, in which they depend only on $X^{(1)}$, with the respective polarization tensors $\epsilon^{a(s)}$ being by construction symmetric, represented by the Young diagram
\begin{align}
    \YY{s}: \quad \textcolor{red}{\RectARow{4}{$s$}}\,.
\end{align}
Generically, such an $F$ can be written as
\begin{align} \label{simplest_principal}
     F_\epsilon^f=f(a^\dag_I)\, F_\epsilon (X^{(1)}) \,,
\end{align}
where $a^\dag_I=\{ T^{01}, T^{11} \}$. However, the Virasoro constraints (\ref{leadingtrajex}) imply that $f=1$, so $h_f=w=0$ and (\ref{simplest_principal}) belongs to the principal embedding (\ref{pol_princ}) and describes a single trajectory, the leading Regge trajectory reviewed in section \ref{sec:leading}, highlighted in red in tables \ref{firstseven} and \ref{depth_traj}. This is the simplest example of the fact that for fixed $K$, we are able to detect only a finite number of trajectories.   

\paragraph{$\boldsymbol{K=2}$.} Next, we consider polynomials that depend only on $X^{(1)}$, $X^{(2)}$, with the respective polarization tensors  $\epsilon^{a(s_1),\, b(s_2)}$ generically having the symmetry of an arbitrary 2--row Young diagram 
\begin{align}\label{tworowstates}
   \YY{s_1,s_2}: \quad \RectBRowUp{6}{4}{$s_1$}{$s_2$}\,.
\end{align}
Generically, such an $F$ can be written as
\begin{align} \label{next_principal}
  F_\epsilon^f=f(a^\dag_I) F_\epsilon (X^{(1)}, X^{(2)}) \,,
\end{align}
where $a^\dag_I=\{ T^{01}, T^{02}, T^{11}, T^{12},T^{22}, T\fud{2}{1} \}$. For $f=1$, the states described by \eqref{next_principal} appear at level $N_{\textrm{min}}=s_1+2s_2$. Restricting to the transverse target space, we reduce the list of creation operators to $a^\dag_I=\{ T^{11}_\perp, T^{12}_\perp,T^{22}_\perp, T\fud{2}{1} \}$ and, save for $(L_0-1)F=0$, the Virasoro constraints \eqref{VirasoroB} simplify further to
\begin{align}
    T\fud{1}{2}F_\epsilon^f= T_{11}F_\epsilon^f=T_{12}F_\epsilon^f=T_{22}F_\epsilon^f=0\,,
\end{align}
which are precisely the lowest--weight conditions for $sp(4)$. They imply $f=1 \Rightarrow w=0$. Consequently, the physical (\ref{next_principal}) covers \textit{all} diagrams at $w=0$ with two rows at most, namely the leading Regge, the trajectories highlighted in blue and violet in table \ref{depth_traj}, as well as (infinitely) many others that have more boxes in the second row. All together they form a family with polarization tensors having symmetries of two-row Young diagrams. To give an example and illustrate the power of the transverse Ansatz, let us consider the simplest (massive) low--spin example of $K=2$, namely
\begin{align}
    F&= \tfrac12 F^{\mu\nu} X^{(1)}_\mu X^{(1)}_\nu + F^\mu X^{(2)}_\mu \,,
\end{align}
which is precisely the level $N=2$ (\ref{open_2}) we have already reviewed in the old--school formulation.  Using \eqref{spurious_new}, BRST--exact states take the form
\begin{align} \label{spurious_new_ex}
    V_{\textrm{sp}} = \big[ip\cdot X^{(1)}+T\fud{2}{1} \big] \, G \, e^{ip\cdot X^{(0)}}  = \frac{i}{\sqrt{2\alpha'}}  \Big[i p\cdot X^{{(1)}} \, \epsilon \cdot X^{(1)}  + \epsilon \cdot X^{(2)} \Big] \, e^{ip\cdot X^{(0)}} \quad , \quad p^2=-\frac{1}{\alpha'} \,,
\end{align}
where
\begin{align}
    G=\frac{i}{\sqrt{2\alpha'}} \epsilon^\mu X^{(1)}_\mu  \,,
\end{align}
matching \eqref{spurious_2}, which are associated with the gauge symmetry \eqref{gauge_N2} that can be used to eliminate $F^\mu$. As a result, we capture a transverse--traceless state from the leading Regge trajectory, the massive spin--2 tensor. On the other hand, applying our considerations for \eqref{next_principal} in the transverse target space, we immediately have
\begin{align}
    N=N_{min}=2 \quad \Rightarrow  \quad s_1=2\,,\, s_2=0
\end{align}
without having to treat BRST--exact states. To summarize, the lightest levels $N=0,1,2,3$ reviewed in Subsection (\ref{sub:lightex}) consist of states that find themselves all at $w=0$.

\paragraph{$\boldsymbol{K=3}$.} Let us also present the lightest member--state of the trajectory highlighted in teal in tables \ref{firstseven} and \ref{depth_traj}, the $(1,1,1)$--tensor at level $N=6$. Its Young symmetry and level fix entirely its polynomial to
\begin{align}
    F&=  F^{\mu\nu \lambda } X^{(1)}_\mu X^{(2)}_\nu X^{(3)}_\lambda \,,
\end{align}
where $F^{\mu\nu \lambda }$ is totally antisymmetric and transverse. Since $w=0$, the dressing function is trivial and following the trajectory along essentially means constructing polarization tensors with the appropriate Young symmetry and at the right level, which simply gives
\begin{align}
    F&=  F^{\mu(s_1), \nu(s_2), \lambda(s_3) } X^{(1)}_{\mu_1}\ldots X^{(1)}_{\mu_{s_1}} \, X^{(2)}_{\nu_1}\ldots X^{(2)}_{\nu_{s_2}}\, X^{(3)}_{\lambda_1}\ldots X^{(3)}_{\lambda_{s_3}} \,.
\end{align}

\paragraph{Non--existent trajectory.} It may also be instructive to search once for non--existent states to test the formalism. For example, one may imagine that there is a spin--$(s-1)$ family at level $N=s$, since we can write
\begin{align} \label{wrong_spin}
    F_\epsilon^f= \epsilon^{a(s-1)}\, \bigg[a_1(s-2)  X^{(1)}_a\ldots X^{(1)}_a X^{(2)}_a+ a_2  X^{(1)}_a\ldots X^{(1)}_a \,\Big(  ip\cdot X^{(1)}\Big) \bigg]\,,
\end{align}
with $a_1$, $a_2$ dimensionless parameters, which can be rewritten as
\begin{align}
    F_\epsilon^f \equiv f \,F_\epsilon  = \big[ a_1 T\fud{2}{1} + ia_2\, T^{01} \big]\, F_\epsilon\,,
\end{align}
where
\begin{align} \label{seed_constr}
 F_\epsilon \equiv \epsilon^{a(s-1)}\, X^{(1)}_a\ldots X^{(1)}_a \equiv \epsilon^{a(s-1)}\, X^{(1)}_{a(s-1)}   \quad , \quad T^0_{\hphantom{0}1} F_\epsilon  =0  \,, \quad   T_{11}\, F_\epsilon=0  \,,
\end{align}
and
\begin{align}
    h_{F_\epsilon^f}=s \quad , \quad h_{F_\epsilon}= s-1 \quad, \quad h_f=w=1\,,
\end{align}
namely $f$ is supposed to uplift the trajectory consisting of spin--$(s-1)$ states at level $s-1$ to level $s$. For $F_\epsilon^f$ to correspond to a physical state, it has to satisfy the conditions \eqref{VirasoroB} since $K=2$, which yield
\begin{align}
  p^2 = -\frac{s-1}{\alpha'} \quad, \quad a_1=a_2\equiv a \quad \Rightarrow \quad F_\epsilon^f=a\,[T^2_{\hphantom{2}1}+ i \, p\cdot X^{(1)}]\, F_\epsilon \,,
\end{align}
namely $F_\epsilon^f$ takes the form of a BRST--exact state (\ref{spurious_new}), the simplest example of which being (\ref{spurious_new_ex}). As expected from the discussion after equation (\ref{mass_covariant_spurious}), the spin--$(s-1)$ state created by $F_\epsilon$ is a spurious state at level $s$ and a physical state at level $s-1$. In the transverse subspace, the second term in \eqref{wrong_spin} is forbidden and the constraints \eqref{VirasoroB} simply tell us that the first one cannot pass.

\subsection{Non--principal embeddings}

We now focus on non--principal embeddings and construct various examples of dressing functions. Since $K=1,2$ inevitably lead to the principal embedding, we have to set at least $K=3$. For every family of trajectories, the input consists of the value of $w$ as well as  of the lengths of the rows of the depicted Young diagrams, which using (\ref{depthdef}) yields $N$.

\paragraph{$\textcolor{olive}{\RectARow{5}{$s-2$}}$--trajectory, $\boldsymbol{w=2}$.} This is the trajectory highlighted in olive in tables \ref{firstseven} and \ref{depth_traj}, with its lightest member being a massive spin--2 tensor at $N=4$. Let us construct its most general Ansatz by uplifting the trajectory of spin--$(s-2)$ states at $N_\textrm{min}= s-2$ of the principal embedding $w=0$ 
\begin{align} \label{seed_2}
    F_\epsilon\equiv \Big( \frac{i}{\sqrt{2\alpha'}}\Big)^{s-2} \epsilon^{a(s-2)}X^{(1)}_{a(s-2)} \quad , \quad T_{11}F_\epsilon=T\fud{0}{1}F_\epsilon=0 \quad , \quad T\fud{1}{1}F_\epsilon=(s-2)F_\epsilon\,,
\end{align}
to $N=s>4$, $w=2$. The Ansatz can span oscillators up to $X^{(3)}$, namely
\begin{equation}
    \begin{array}{cll}
     \label{subleading_first}
    F_\epsilon^f&= \Big( \frac{i}{\sqrt{2\alpha'}}\Big)^{s-2} \, \epsilon^{a(s-2)}\bigg[\frac{\beta_1}{\alpha'}\, X^{(1)}_{a(s-2)} (X^{(1)}\cdot X^{(1)})+ \beta_2\, X^{(1)}_{a(s-2)}(ip\cdot X^{(1)})^2 
    \crbig
    & \qquad  +\beta_3 \, X^{(1)}_{a(s-2)} (ip\cdot X^{(2)})  + (s-2) \,\beta_4 \, X^{(1)}_{a(s-3)} X^{(2)}_a(ip\cdot X^{(1)}) 
    \crbig
    & \qquad + (s-2)\,\beta_5 \, X^{(1)}_{a(s-3)}X^{(3)}_a  + (s-2)(s-3) \, \beta_6 X^{(1)}_{a(s-4)}X^{(2)}_{a}X^{(2)}_a  \bigg]\,,
    \end{array}
\end{equation}
where the parameters $\beta_1 , \ldots , \beta_6$ are real and dimensionless, which can be rewritten as  
\begin{align} \label{subleading_first_c}
\begin{aligned}
        F_\epsilon^f\equiv f F_\epsilon & = \bigg[ \frac{\beta_1}{\alpha'}T^{11} + \beta_2(ip\cdot X^{(1)} )^2 +\beta_3(ip\cdot X^{(2)})     + \beta_4(ip\cdot X^{(1)}) T\fud{2}{1}
    \crbig
    & \qquad \qquad + \beta_5 T\fud{3}{1}+\beta_6(T\fud{2}{1})^2 \bigg] F_\epsilon\,,
    \end{aligned}
\end{align}
with the weights
\begin{align}
    h_{F_\epsilon^f}=s \quad , \quad h_{F_\epsilon}= s-2 \quad, \quad h_f=w=2\,,
\end{align}
fixing its structure. The Virasoro constraints (\ref{simplest_eq_3_new_not_1})--(\ref{simplest_eq_3_new_not_3}) for $K=3$ on (\ref{subleading_first_c}) lead to the on--shell condition
\begin{align} \label{onshell_sublead}
    p^2 = -\frac{s-1}{\alpha'}
\end{align}
as well as to three algebraic relations involving the $\beta$'s, which we can use to write three parameters in terms of $\beta_1$, $\beta_3$, $\beta_4$ and $s$. It is convenient to perform the shift
\begin{align}\label{shift}
s-2 \rightarrow s\,,    
\end{align}
so that all member--states of the trajectory have spin--$s$ and appear at $N=s+2$. The solution reads
\begin{align} \label{solution_w2_1}
    \beta_2&=-\frac{\beta _1}{s+1}+\frac{\beta _3}{2 (s+1)}+\frac{\beta _4 s}{2 (s+1)}\,, \\ \label{solution_w2_2}
    \beta_5&=\frac{\beta _1 (d+2 s-1)}{3 s}+\frac{\beta _3 (4 s+5)}{6 s}+ \frac{\beta _4}{6}\,,\\ 
    \beta_6&=-\frac{\beta _1 (d+2 s-1)}{2 (s-1) s}-\frac{\beta _3 (4 s+5)}{4 (s-1) s} +\frac{\beta _4 (2 s+1)}{4 (s-1)}\,. \label{solution_w2_3}
\end{align}

Forgetting for a moment the shift \eqref{shift}, let us construct the corresponding BRST--exact states. Those due to $L_{-1}$ can be written as vertex operators (\ref{arbitrary_op}) of weight $0$ satisfying now the on--shell condition (\ref{onshell_sublead}). The respective $G$ can span oscillators up to $X^{(2)}$ namely takes the generic form 
\begin{align}
    G_\epsilon^{\widetilde{f}}&= \Big( \frac{i}{\sqrt{2\alpha'}}\Big)^{s-2} \, \epsilon^{a(s-2)}\Big[ \gamma_1 X^{(1)}_{a(s-2)} (ip\cdot X^{(1)}) + (s-2) \,\gamma_2 X^{(1)}_{a(s-3)} X^{(2)}_a\Big]
\end{align}
or
\begin{align}
    G_\epsilon^{\widetilde{f}} \equiv \widetilde{f} F_\epsilon= \big[ \gamma_1 (ip\cdot X^{(1)})  + \gamma_2 T\fud{2}{1} \big] F_\epsilon \,,
\end{align}
where the parameters $\gamma_1, \gamma_2$ are real and dimensionless and the weights
\begin{align}
 h_{G_\epsilon^{\widetilde{f}}}=s-1 \quad , \quad h_{F_\epsilon}= s-2 \quad, \quad h_{\widetilde f}=1\,.
\end{align}
have fixed the structure of the Ansatz. $G_\epsilon^{\widetilde{f}}$ further has to satisfy the conditions \eqref{VirasoroB}, which yield
\begin{align} \label{gamma_coeff}
    \gamma_1 = \frac{s-2}{s-1} \gamma_2 \,.
\end{align}
The respective spurious states of spin--$(s-2)$ are then given by
\begin{align}
\begin{aligned}
    F_{\textrm{sp}} &= D\, G_\epsilon^{\widetilde{f}} =\Big[  \gamma_1  (ip\cdot X^{(1)} )^2   + \gamma_1 (ip\cdot X^{(2)})  + \gamma_2 T\fud{2}{1}T\fud{2}{1}
    \\
    & \qquad  \qquad  \qquad + (\gamma_1 + \gamma_2) (ip\cdot X^{(1)}) T\fud{2}{1}+ \gamma_2 T\fud{3}{1} \Big] F_\epsilon \,,
\end{aligned}
\end{align}
with $\gamma_1$ given by (\ref{gamma_coeff}). This allows one to eliminate one free parameter. Another one can be eliminated by employing the $L_{-2}$ generator. Therefore, the physical trajectory in question is parametrized by a single parameter and all its member--states have multiplicity one. 

It is much simpler to construct the same trajectory in the transverse subspace. Here, the Ansatz simplifies to 
\begin{align} \label{subleading_first_cTr}
    F_\epsilon^f&= \bigg[ \frac{\beta_1}{\alpha'}T^{11}_\perp + \beta_5 T\fud{3}{1}+\beta_6(T\fud{2}{1})^2 \bigg] F_\epsilon\,.
\end{align}
Equivalently, the parameters $\beta_{3,4}$ can be eliminated by BRST--exact terms (one needs to use both $D\sim L_{-1}$ and $L_{-2}$, thereby fixing the ambiguity, while the transverse Ansatz requires a specific relation between $\beta_{1,2}$). The solution reads
\begin{align} \label{solutiontr} 
    \beta_5= \beta_1 \frac{d+2 s-1}{3s} \quad , \quad \beta_6=-\beta_1 \frac{d+2 s-1}{2(s-1)}\,,
\end{align}
where we have performed the shift \eqref{shift} and the relation between $\beta_{1,2}$ is consistent with having $T^{11}_\perp$ as well. The singularities of the coefficients at $s=0,1$ reveal that indeed the trajectory starts with $s=2$ at $N=4$ as expected from table \ref{firstseven}, whose vertex operator we can now easily write by setting $s=2$, $d=26$ in the trajectory's solution
\begin{equation} \label{massivegr2}
    \begin{array}{ccl}
  V_{\widetilde B}(p,z)=  \frac{1}{2\alpha'} \, \widetilde{B}_{\mu \nu}  \, \Big[  -\frac{3}{\alpha'}\, i\partial X^{\mu}  i\partial X^{\nu}  i \pl X^\kappa i \pl X_\kappa   + 29 \, i\partial X^{\mu}  i\partial^3 X^{\nu} -87 \, i\partial^2 X^{\mu}  i\partial^2 X^{\nu}\, \Big] e^{ip \cdot X} 
    \end{array}\,.
\end{equation}
This is the second lightest massive spin--$2$ state that appears in the open bosonic string spectrum and it is the lightest state at a non--trivial depth that our formalism probes.

As this first example of non--principal embedding shows, spin is a free parameter here (boxes can be added to the first row of the Young diagram). This demonstrates the efficiency of the depth--by--depth approach compared to proceeding level--by--level: writing the Ansatz for the level $N=4$ and solving the Virasoro constraints to determine the physical states can certainly give access to (\ref{massivegr2}), but will not provide information about other levels. Instead, the solution \eqref{solutiontr} covers the vertex operators of the entire trajectory with lowest member the state (\ref{massivegr2}). Notice also that we cannot add boxes to the second row since it was empty. If we took $\epsilon$ to enjoy the symmetry of a two--row Young diagram, this would activate additional terms in $L_n$ and our dressing function would fail. More generally, the complexity of the dressing function is due to the number of rows as well as the depth $w$ of the trajectory we want to construct.  

From now on we consider dressing functions in the transverse subspace, which simplifies their form and also completely fixes the ambiguity due to BRST--exact terms.  

\paragraph{$\RectBRow{7}{3}{$s_1-2s_2-1$}{$s_2$}$\,--trajectories, $\boldsymbol{w=1}$.} This is a double--infinite family of states that, for any admissible $s_2$, appears at level $N=s_1$: the two trajectories displayed at $w=1$ in table \ref{depth_traj} are precisely of this type. If this family were to be principally embedded, it would appear at $N=s_1-1$, which case we have already treated. Now, the principal embedding is just by one unit of energy away. Therefore, the dressing function has to be very simple; indeed, in the transverse subspace the only operators of weight $1$ are $T\fud{k+1}{k}$, $k=1,2,\ldots\,$. It is more convenient to reparameterize the family as $\RectBRowUp{5}{3}{$s_1$}{$s_2$}$, rewriting its level as $N=s_1+2s_2+1$ and the polynomial to be uplifted depends on $X^{(2)}$ at most. Then the Ansatz for the dressing function reads
\begin{align}\label{oneunitA}
    f&= T\fud{2}{1}+a T\fud{3}{2}\,.
\end{align}
The solution of the Virasoro constraints for $K=3$ is $a=-(s_1-s_2)/(3s_2)$. This example illustrates another general feature of the formalism: the various lengths of the rows of Young diagrams can be kept arbitrary, which allows to excavate \textit{multiple} trajectories at a time, i.e. families. 

\paragraph{$\RectCRow{6}{5}{3}{$s_1$}{$s_2$}{$s_3$}$\,--trajectories and beyond, $\boldsymbol{w=1}$.} At very low cost, we can find similar states with more complicated Young diagrams that appear at level $N = s_1+2s_2+3s_3+1$ and beyond and are all one unit of energy away from the principal embedding. The dressing function is the same \eqref{oneunitA}, but the solution is $a=-(s_1-s_2)/(3(s_2-s_3))$ and works for any number of rows.

There are also other states that are one unit away from the principal ones and have the same spin. For example, one can shift the indices of the dressing function \eqref{oneunitA} by one to get
\begin{align}
    f&= T\fud{3}{2}+a T\fud{4}{3}\,,
\end{align}
and find that $a=-(s_2-s_3)/(2(s_3-s_4))$ using the Virasoro constraints for $K=4$. Even more generally, whenever the corresponding Young diagram makes sense in $d-1$ (transverse) dimensions, we can use
\begin{align}
    f_i= T\fud{i}{i-1}+a T\fud{i+1}{i} \quad , \quad  a=-\frac{(i-1)(s_{i}-s_{i-1})}{(i+1)(s_{i+1}-s_i)}\,.
\end{align}
Note that for all these trajectories, the $L_2$ condition is automatically satisfied: $T\fud{i+1}{i}$ destroy Young symmetry between the rows $i$ and $i+1$, but $L_2\sim T\fud{m+1}{m+3}$ checks it across two rows, which is not violated. We can also try a weight $1$ dressing function that acts on rows $i$ and $j$
\begin{align}
    f&= T\fud{i+1}{i}+a T\fud{j+1}{j}\,.
\end{align}
The $L_1$ condition implies 
\begin{align}
     a=-\frac{i(i+1)(s_{i+1}-s_{i})}{j(j+1)(s_{j+1}-s_j)}\,.
\end{align}
The last example covers \textit{all} possible trajectories at $w=1$. We observe that the set of principally embedded trajectories represented by Young diagrams with $n$ rows can be shifted in $n-1$ different ways one level up, which determines the multiplicity. 

\paragraph{$3^{*}\,\RectBRow{7}{3}{$s_1-2s_2-2$, }{$s_2$}$\,--trajectories, $\boldsymbol{w=2}$.} This is a triply degenerate family of trajectories; the second trajectory displayed at $w=2$ in table \ref{depth_traj} is of this type. It is again convenient to reparameterize the family as $\RectBRowUp{5}{3}{$s_1$}{$s_2$}$, such that it appears at level $N=s_1+2s_2+2$. Then the Ansatz for the dressing function reads
\begin{align}
    f&= \frac{\beta_1}{\alpha'}T^{11}_\perp +\beta_2(T\fud{2}{1})^2 +\beta_3 T\fud{2}{1}T\fud{3}{2} +\beta_4 T\fud{3}{2}T\fud{3}{2} + \beta_5 T\fud{3}{1}+\beta_6 T\fud{4}{2}\,,
\end{align}
so $K=4$. There are three free parameters in the solution of the respective Virasoro constraints. Three independent solutions can be chosen to be
\begin{align}
    \beta_i&=\left\{\frac{6}{d+2 s_1-1},-\frac{3}{s_1 \left(s_1-s_2-1\right)},0,\frac{1}{3 s_1 \left(s_2-1\right)},\frac{2}{s_1},0\right\}\,,\\
    \beta_i&= \left\{0,-\frac{3 \left(s_1+1\right) s_2}{2 \left(s_1-s_2-1\right)},s_1,-\frac{\left(s_1+1\right) \left(s_1-s_2\right)}{6 \left(s_2-1\right)},s_2,0\right\}\,,\\
    \beta_i&=\left\{0,\frac{6 s_2}{s_1-s_2-1},0,-\frac{3 s_1+2 s_2}{3 \left(s_2-1\right)},-4 s_2,s_1\right\}\,.
\end{align}
The first solution reduces to \eqref{subleading_first_cTr} for $s_2=0$.

\paragraph{$\RectARow{5}{$s-3$}$--trajectory, $\boldsymbol{w=3}$.} This is a trajectory of symmetric states deeper into the spectrum. The Ansatz for the dressing function reads
\begin{align}
    f&= \frac{\beta_1}{\alpha'} T^{12} +\beta_2 T\fud{4}{1}+\beta_3 T\fud{2}{1}T\fud{2}{1}T\fud{2}{1}+\frac{\beta_4}{\alpha'} T^{11}T\fud{2}{1}+\beta_5 T\fud{3}{1}T\fud{2}{1} \,.
\end{align}
The unique solution (up to an overall factor) of Virasoro constraints for $K=4$ can be chosen to be 
\begin{align}
    \beta_i &= \left\{\frac{s}{d+s-3},\frac{1}{6},\frac{1}{(s-2) (s-1)},-\frac{1}{d+s-3},-\frac{1}{s-1}\right\}\,,
\end{align}
where again we have shifted $s-3 \rightarrow s$, so that the family appears at level $N=s+3$.

\paragraph{$3^{*}\,\RectARow{5}{$s-4$}$--trajectory, $\boldsymbol{w=4}$.} Even deeper, the Ansatz for the dressing function reads 
\begin{align}
\begin{aligned}
    f&=\frac{\beta_1}{\alpha'}T^{22} +\frac{\beta_2}{\alpha'} T^{13} +  \beta_3T\fud{5}{1} +\beta_4(T\fud{2}{1})^4  +\frac{\beta_5}{\alpha'}T^{1,1}(T\fud{2}{1})^2 +\beta_6 T\fud{3}{1} (T\fud{2}{1})^2 +\frac{\beta_7}{\alpha'}T^{12} T\fud{2}{1}+ \\   
    &\qquad \beta_8 T\fud{4}{1} T\fud{2}{1}+\frac{\beta_9}{\alpha'^2}(T^{11})^2  + \frac{\beta_{10}}{\alpha'}T^{11} T\fud{3}{1}   + \beta_{11} (T\fud{3}{1})^2 \,.
\end{aligned}
\end{align}
The Virasoro constraints for $K=5$ have three independent solutions. 

\paragraph{Complexity of dressing functions.} The complexity of dressing functions depends on the height of Young diagrams $n$ we want to search for and on the depth $w$ of the trajectory. Indeed, the most general Ansatz of a given weight $w$ is built from $T^{kl}$, $k+l\leq w$, $k,l\leq n$ and $T\fud{k}{l}$, $0< k-l\leq w$, $l\leq n$. We could easily cover all cases of $w=1$ and an arbitrary Young diagram. Note that the $sp(2\bullet)$--module generated by creation operators $a^\dag_I$ is infinite dimensional (it is finite for any given tensor, but the dimension depends on the Young diagram and can grow indefinitely). Therefore, if we restrict the number of oscillators $\alpha_{-n}$ to $n\leq K$ and want to search for all trajectories (i.e. all depths $w$), with such a restriction the Ansatz is infinite. For example, for the leading trajectory in the transverse subspace the dressing function could have been an arbitrary function of $T^{11}$. However, the Virasoro constraints \eqref{leadingtrajex} leave $f=1$ as the unique solution. In addition, we know that for fixed $K$ the number of trajectories is finite and, hence, there is a way to improve the formalism.  

To summarize, all trajectories displayed in table \ref{depth_traj} have been analyzed in this Subsection. It is also straightforward to proceed to $w=5$ and $w=6$ which can probe the vector and singlet of $N=6$. All other states of table \ref{firstseven} have been covered.

\paragraph{Closed strings.} In order to treat the case of closed string states, a few modifications are needed. The modes $X^{(k)}_\mu=\pl^{(k)} X_\mu$ need to be extended with $\ov{X}^{(k)}_\mu=\bar{\pl}^{(k)} X_\mu$ and we can use our open string dressing functions to write functions that depend on $X^{(k)}_\mu$, for the left movers, and on $\ov{X}^{(k)}_\mu$, for the right movers. The total dressing function of closed string is just the product of any two of the open provided the level matching condition is imposed, $L_0=\bar{L}_0$.

\section{Tree--level amplitudes}
\label{sec:amplitudes}

In this Section, we compute tree--level amplitudes for the scattering of open strings, specializing in the ``deeper'' trajectories constructed in Section \ref{sec:slices}. An open string tree--level amplitude is the Wick contraction of the (normal--ordered) vertex operators creating the open string external states at locations $x$ of the boundary $\mathbb R$ of the disk $\mathbb{D}_2$, which is summed over the inequivalent orderings $\sigma$ of, as well as integrated over, the insertion points. The contribution of the volume of the conformal Killing group of $\mathbb{D}_2$ is canceled after inserting a $c$--ghost at every open string insertion and fixing the locations of three (arbitrary) vertex operators. For more details, see for example \cite{Polchinski:1998rq, Blumenhagen:2013fgp}. Normalizing our amplitudes is straightforward, so we suppress overall normalization factors.

\subsection{N--point technicalities}

The simplest $N$--point function to compute is that of the open string tachyon \eqref{open_tachyon}, which reads
\begin{align} \label{KN_open}
  \mathcal{E}_N \equiv   \left\langle :V_{\textrm{tachyon}}^{\textrm{open}}(p_1,z_1):\, \ldots : V_{\textrm{tachyon}}^{\textrm{open}}(p_N,z_N): \right\rangle = \left \langle \displaystyle\prod_{i=1}^{N}:e^{ip_i\cdot X_i}: \right \rangle_{\mathbb{D}_2}   \,.
\end{align}
Using the open string two--point function (\ref{boson_corr}), (\ref{KN_open}) gives the universal Koba--Nielsen factor for open string amplitudes
\begin{align}
\mathcal{E}_N =    \prod_{i<j}^N |z_{ij}|^{2\alpha' \, p_i \cdot p_j}\,.
\end{align}
Momentum conservation  $\sum_i^N p_i^\mu = 0$ is always implicit. We will be focusing on the trajectory of symmetric tensors of spin--$s$ at level $N=s+2$ (which we originally treated in the form spin--$(s-2)$ at $N=s$) and $w=2$ that is highlighted in olive in tables \ref{firstseven} and \ref{depth_traj}, namely the one described by vertex operators built by the dressing functions \eqref{subleading_first_c} at the solution \eqref{solution_w2_1}--\eqref{solution_w2_3}, or, in transverse space, by \eqref{subleading_first_cTr} at the solution \eqref{solutiontr}. For convenience, we will be referring to this family of states as the $w=2$ trajectory. Let us highlight that the form of the on--shell condition for the leading Regge as well as for the $w=2$ trajectory, is the same
\begin{align}
p_i^2 = -\frac{l_i-1}{\alpha'}    \,,
\end{align}
where $l_i$ is the level of the state of the $i$--th leg.

For the leading Regge trajectory, the generating function \eqref{gen_xi} can be used, the $N$--point function for which reads \cite{Kawai:1985xq}, see also \cite{Gross:1986mw, Sagnotti:2010at}, 
\begin{align} 
\mathcal{Z}_N^{\textrm{leading}} = \left \langle \displaystyle \prod_{i=1}^{N} : \exp \big( i\, p_i \cdot X_i  + i \, \xi_i \cdot \partial X_i \big): \right \rangle_{\mathbb{D}_2}  
=\mathcal{E}_N\, \exp \bigg[ \displaystyle \sum_{i\neq j}^N \alpha' \Big(  2  \frac{\xi_i \cdot p_j }{z_{ij}} + \frac{\xi_i \cdot \xi_j}{z_{ij}^2} \Big) \bigg] \,. \label{Z_leading}
\end{align}
For more general trajectories, it is convenient to start with a simple generating function
\begin{align}
   \mathbf V\big(z,p,\vec{\xi}\,\big)&=\exp \big( i\, p \cdot X +i \, \xi^{(1)} \cdot \partial X+i \, \xi^{(2)} \cdot \partial^2 X+\ldots \big)\equiv \exp \bigg( i\, \sum_{n=0}^\infty \xi^{(n)} \cdot X^{(n)}\bigg)\,,
\end{align}
where $\xi^{(0)}\equiv p$ and the auxiliary vectors $\xi^{(n)}_\mu$ obey initially no relations. We then find the general $N$--point correlation function to be
\begin{align}
\begin{aligned}
    \mathcal{Z}_N & = \left \langle \displaystyle \prod_{i=1}^{N} : \exp \bigg( i\, \sum_{n_i=0}^\infty \xi_i^{(n_i)} \cdot X_i^{(n_i)}\bigg): \right \rangle_{\mathbb{D}_2}  
    \crbig
    &\quad= \mathcal{E}_N\exp \bigg[ \sum_{i\neq j} \sum_{ n>0}(-)^{n+1} \alpha' \bigg( 2  (n-1)! \frac{\xi_i^{(n)}  \cdot p_j}{z_{ij}^{n}}  +\sum_{m>0} (n+m-1)! \frac{\xi_i^{(n)}\cdot \xi_j^{(m)} }{z_{ij}^{n+m}}\bigg) \bigg]\,, \label{Z_any}
\end{aligned}
\end{align}
where we recall that three kinds of indices are distinguished, namely spacetime indices $\mu,\nu, \ldots\,$, lower indices $i,j,\ldots\,$ that denote worldsheet location / leg number and upper indices $(n),(m)$ that denote components of the (Euclidean and infinite--dimensional) vectors $\vec{\xi}^\mu$ and $\vec{X}^\mu$\,.

Now, suppose we are interested in a family of states with polarization tensors having the symmetries of $n$--row Young diagrams $\YY{s_1,\ldots,s_n}$. It is convenient to factorize the polarization tensor $\epsilon^{a(s_1),\ldots, c(s_n)}$ into the products of polarization vectors $\epsilon^a_I$, $I=1,\ldots, n$. Tracelessness and transversality are then enforced by the additional constraints 
\begin{align} \label{fact_pol}
    \epsilon_I \cdot \epsilon_J =0 \quad , \quad  \epsilon_I \cdot p=0 \,.
\end{align}
Let us denote the $\YY{s_1,\ldots,s_n}_{\epsilon_1,\ldots, \epsilon_n}$ the Young symmetrizer that projects onto the indicated Young symmetry the tensor product of $\epsilon_I$'s. One possible realization of such a projector was given in \eqref{YoungProjector}. Then trajectories belonging to the principal embedding, $w=0$, can be obtained via
\begin{align}
   \YY{s_1,\ldots,s_n}_{\epsilon_1,\ldots, \epsilon_n} \prod_{k=1}^{k=n}(-i\epsilon_k\cdot \pl_{\xi^{(k)}})^{s_k}  \mathbf V\big(z,p,\vec{\xi}\,\big)\big|_{\xi^{(i)}=0}\,.
\end{align}
For example, for the leading Regge trajectory \eqref{ansatz_leading}, we have 
\begin{align}
   \mathbb V_{F_1}\big(z,p,\xi^{(1)}=-i \epsilon^{(1)},0,\ldots,0\,\big)=  \exp[-i \epsilon^{(1)}\cdot \pl_{\xi^{(1)}}]\, \mathbf V\big(z,p,\vec{\xi}\,\big) \big|_{\xi^{(i)}=0}\,,
\end{align}
since the Young projector trivializes, which is essentially \eqref{gen_xi}. Turning to $w>0$, any dressing function $f$ can be represented as a poly--differential operator in $\xi^{(n)}$ that acts on $\mathbf V\big(z,p,\vec{\xi}\,\big)$. Recalling that a generic $f$ contains a number of derivatives in $\partial X$, together with the coefficients in front of its corresponding terms, one finds the following set of functions
\begin{align}
    E_k(x)&= \sum_{s=k} x^{s-k}/s! \,,\, && E_0(x)=e^x \,,\, && E_1(x)=\tfrac{e^x-1}{x} \,,\, && E_2(x)=\tfrac{e^x-1-x}{x^2}\,,
\end{align}
which are truncated exponents. $E_k(x)$ can be massaged back into a purely exponential form with the help of 
\begin{align}
    E_k(x)&= \int_{\Delta_k} \exp[ u_1 x]\,,
\end{align}
where $\Delta_k=\{0\leq u_1 \leq \ldots \leq u_k\leq 1\}$ is the $k$--dimensional simplex. For example, the $w=2$ trajectory \eqref{subleading_first_cTr} can be represented as
\begin{align}\label{subsimple}
\begin{aligned}
    \mathbb V&= \int_{\Delta_2} e^{-i u_1\epsilon \cdot \pl_{\xi^{(1)}}} \bigg\{ \delta(1-u_2)\delta(u_2-u_1) \Big[-(\pl_{\xi^{(1)}}\cdot_\perp\pl_{\xi^{(1)}}) -\tfrac23 i \epsilon \cdot\pl_{\xi^{(3)}} \Big]\\
    &\qquad \qquad +\delta(u_2-u_1) \Big[\tfrac{d-1}{3}(-i \epsilon \cdot \pl_{\xi^{(3)}})+(\epsilon \cdot \pl_{\xi^{(2)}})^2 \Big] +\tfrac{d-1}{2}(\epsilon \cdot \pl_{\xi^{(2)}})^2 \bigg\} \, \mathbf V\big(z,p,\vec{\xi}\,\big) \big|_{\xi^{(i)}=0}\,.
\end{aligned}
\end{align}
In general, after applying the derivatives that are inside $f$, one needs to replace the corresponding $\xi$ with $-i u_1 \epsilon$. 

For completeness, let us note that for closed string amplitudes, we have to first double the generating function
\begin{align}
   \mathbf  V_{\vec{\xi},z,\bar{z}}&= \exp \bigg( i\, \sum_{n=0}^\infty \xi^{(n)} \cdot X^{(n)}(z)+i\, \sum_{n=0}^\infty \xi^{(-n)} \cdot \ov X^{(n)}(\bar z)\bigg)\,,
\end{align}
and then generalize the open string considerations.

\subsection{3--point examples}
What is special about $3$--point amplitudes is that (i) the kinematic structure is very simple and for fixed spins there is always a finite number of independent structures; (ii) momentum conservation fixes the scalar products of all momenta to be
\begin{align}
    2 \alpha' \, p_i \cdot p_j = l_i +l_j -l_k-1 \quad , \quad i \neq j \neq k\,.
\end{align}
Therefore, the product of the Koba--Nielsen factor at $3$ points and the $c$--ghost contribution turns into
\begin{align} \label{KN_ghost}
\mathcal{E}_3 \times     \langle c_1 c_2 c_3 \rangle=    |z_{12}|^{l_1+l_2-l_3} \, |z_{13}|^{l_1+l_3-l_2} \, |z_{23}|^{l_2+l_3-l_1}\,.
\end{align}
For example, if one leading Regge or $w=2$ trajectory and two tachyons are being scattered, \ref{KN_ghost} yields 
\begin{align}
\mathcal{E}_3 \times     \langle c_1 c_2 c_3 \rangle =    \bigg|\frac{z_{12} z_{13}}{ z_{23}}\bigg|^s\,.
\end{align}
It is not immediately obvious that $3$--point amplitudes do not depend on $z_{ij}$, as should be the case. We will be using the notation $(w_1\ldots w_k)$--amplitude to indicate that the $i$--th leg is a trajectory at depth $w_i$, while $\mathcal A^{s_1 s_2 s_3}$ signifies that the $i$--th leg has spin--$s_i$.

\paragraph{Principal$\boldsymbol{{}^{3}}$ or ($\boldsymbol{000}$)--amplitudes.} The case of the $3$--tachyon amplitude is trivial:
\begin{align} \label{KN_ghost_three}
\mathcal{E}_3 \times     \langle c_1 c_2 c_3 \rangle= 1  \quad \Rightarrow \quad    \mathcal{A}^{000} = \Tr(T^{a_1} T^{a_2} T^{a_3} + T^{a_1} T^{a_3} T^{a_2})\,.
\end{align}
The amplitude of one leading leg and two tachyons takes the form 
\begin{align} 
\begin{aligned}
    \mathcal{A}^{s00}&= \langle c_1\,\mathbb V^{a_1}_{F^{\textrm{leading}}} (z_1,p_1)  \, c_2\, V^{a_2}_{\textrm{tachyon}}(z_2,p_2)\, c_3\, V^{a_3}_{\textrm{tachyon}} (z_3,p_3) \rangle  + (p_2,a_2) \leftrightarrow (p_3,a_3)
    \crbig
    &=  \sum_s \bigg(\frac{1}{\sqrt{2\alpha'}}\bigg)^s \,  \bigg|\frac{z_{12}z_{13}}{z_{23}}  \bigg|^s \, \frac{1}{s!}  \bigg(\epsilon \cdot \frac{\pl}{\pl \xi}\bigg)^s \exp\Big\{2\alpha' \xi\cdot p_2 \, \big[z_{12}^{-1} -z_{13}^{-1}\big] \Big\} \bigg|_{\xi=0}
    \crbig
    & \qquad \times \Big[\Tr(T^{a_1} T^{a_2} T^{a_3}) +(-)^s \Tr(T^{a_1} T^{a_3} T^{a_2}) \Big] 
    \crbig
& =\sum_s  (\sqrt{2\alpha'})^s \,\frac{(\epsilon 
\cdot p_2)^s}{s!}  \, \Big[\Tr(T^{a_1} T^{a_2} T^{a_3}) +(-)^s \Tr(T^{a_1} T^{a_3} T^{a_2}) \Big] \,,\label{leading_t_t}
\end{aligned}
\end{align}
where in the first (and second) line we sum over the two possible orderings of the three vertex operator insertions and in the second line we have used momentum conservation along with the constraints (\ref{fact_pol}) for the factorized polarization $F^{\mu_1 \ldots \mu_s}=\epsilon^{\mu_1} \ldots \epsilon^{\mu_s}$ of the leading leg. It is straightforward to see that (\ref{leading_t_t}) matches the result for the same amplitude obtained in \cite{Sagnotti:2010at}; notice that it has no $z$--dependence as appropriate. In the rest of this Subsection, we will be displaying the color--ordered versions of the amplitudes.

The case of three arbitrary spins of the leading Regge trajectory is also well--known. The essential part of the amplitude is \eqref{Z_leading}, the contributing part of the generating exponential of which reduces to
\begin{align*} 
     \exp\bigg\{2\alpha' \bigg[\xi_1\cdot p_2 (z_{12}^{-1}-z_{13}^{-1}) +\xi_2\cdot p_3(z_{23}^{-1}-z_{21}^{-1}) +\xi_3\cdot p_1 (z_{31}^{-1}-z_{32}^{-1}) + \frac{\xi_1\cdot \xi_2} {z_{12}^{2}}+ \frac{\xi_1\cdot \xi_3}{ z_{13}^{2}}+ \frac{\xi_2\cdot \xi_3} {z_{23}^{2}}\bigg] \bigg\}\,,
\end{align*}
upon employing momentum conservation and the constraints (\ref{fact_pol}) for the three legs. Upon expanding this result and collecting terms with fixed powers of the polarization vectors, one observes that its $z$--dependence is canceled by the Koba--Nielsen factor and the $c$--ghost contribution (\ref{KN_ghost_three}). Consequently, one can set for example $z_1=0$, $z_2=-1$, $z_3=1$ to facilitate the computation of $3$--point amplitudes. To give some examples, the amplitude of three (leading) gauge bosons reads\footnote{By $\text{cyclic}[F]$ we mean the sum over all cyclic permutations divided by the number thereof. }
\begin{align}
    \text{cyclic}\Big[2 {\alpha'}  (p_1\cdot \epsilon _3) (p_2\cdot \epsilon _1) (p_3\cdot \epsilon _2)+3 (\epsilon _1\cdot \epsilon _2) (p_1\cdot \epsilon _3)\Big]\,,
\end{align}
which is of course gauge invariant. The amplitude of three leading massive spin--$2$ tensors is given by
\begin{align}
\begin{aligned}
     \text{cyclic}\Big [&2 {\alpha'} ^3 \left(p_1\cdot \epsilon _3\right){}^2 \left(p_2\cdot \epsilon _1\right){}^2 \left(p_3\cdot \epsilon _2\right){}^2+12 {\alpha'} ^2 (\epsilon _1\cdot \epsilon _2) \left(p_1\cdot \epsilon _3\right){}^2 (p_2\cdot \epsilon _1) (p_3\cdot \epsilon _2)\\&+3 {\alpha'}  \left(\epsilon _1\cdot \epsilon _3\right){}^2 \left(p_3\cdot \epsilon _2\right){}^2
     +12 {\alpha'}  (\epsilon _1\cdot \epsilon _2) (\epsilon _1\cdot \epsilon _3) (p_1\cdot \epsilon _3) (p_3\cdot \epsilon _2)\\&+2 (\epsilon _1\cdot \epsilon _2) (\epsilon _1\cdot \epsilon _3) (\epsilon _2\cdot \epsilon _3)\Big]\,,
\end{aligned}
\end{align}
which in a slightly different form can also be found in \cite{Lust:2021jps}. 

A slightly less trivial case is to scatter trajectories of the principal embedding beyond the leading Regge, for example polarization tensors that have the shape of a $2$--row Young diagram, such as the trajectories highlighted in blue and violet in tables \ref{firstseven} and \ref{depth_traj}. Since $w=0$, one does not have to use more than $\mathbf V\big(z,p,\vec{\xi}\,\big)$, e.g. 
\begin{align}
    \mathbb V = \YY{s_1,s_2}_{\epsilon^{(1)},\epsilon^{(2)}}\exp[-i \epsilon^{(1)}\cdot \pl_{\xi^{(1)}}-i \epsilon^{(2)}\cdot \pl_{\xi^{(2)}}]\, \mathbf V\big(z,p,\vec{\xi}\,\big) \big|_{\vec\xi=0}\,,
\end{align}
where the prefactor indicates that the symmetry of a $2$--row Young diagram has to be imposed on $\epsilon^{(1,2)}$. Let's consider the lightest member--state, with diagram $\YoungpAA\,$ ($l=3$), of the blue trajectory, for example. Its amplitude with two tachyons must vanish: it is proportional to $(\epsilon^{(1)}\cdot p_2)(\epsilon^{(2)}\cdot p_2)$, which vanishes after we anti-symmetrize the $\epsilon$'s. Its amplitude with one tachyon and one gauge boson is the first nontrivial one and is proportional to
\begin{align}
    (\epsilon^{(1)}_1\cdot \epsilon^{(1)}_2) (\epsilon^{(2)}_1\cdot p_2)-(\epsilon^{(2)}_1\cdot \epsilon^{(1)}_2) (\epsilon^{(1)}_1\cdot p_2)\,.
\end{align}
To give another example, its amplitude with two gauge bosons is proportional to 
\begin{align*}
    -\alpha '(p_1\cdot \epsilon _3)(p_2\cdot \epsilon _1^{(1)})(\epsilon _1^{(2)}\cdot \epsilon _2)+\alpha '(p_2\cdot \epsilon _1^{(1)})(p_3\cdot \epsilon _2)(\epsilon _1^{(2)}\cdot \epsilon _3)+(\epsilon _1^{(1)}\cdot \epsilon _2) (\epsilon _1^{(2)}\cdot \epsilon _3)- (\epsilon_1^{(1)} \leftrightarrow \epsilon^{(1)}_2)
\end{align*}
Both structures are gauge invariant with respect to the spin--$1$. Note that the on--shell conditions, including Young symmetry, need to be used to show that it's $z$--independent. Amplitudes of tensors with the symmetry of more general Young diagrams are discussed in Appendix \ref{app:scYD}.

\paragraph{($\boldsymbol{200}$)--amplitudes.} The amplitude of one $w=2$ trajectory and two tachyons takes the form
\begin{align}
    \mathcal{A}^{s00}&=\frac{\alpha '^s\left( p_2 \cdot \epsilon  \right){}^s}{3\, s!}q(s) \,,
\end{align}
where 
\begin{align}
    q(s)&= - s\beta _4+{2 \beta _1 (-d+4 s+31)}+{ \beta _3 (2 s+1)}\,.
\end{align}
One can also check that $q(s)$ is BRST invariant. To get a more compact form we restrict to the transverse subspace, which effectively means $\beta_3=\beta_4=0$, 
\begin{align}
    \mathcal{A}^{s00}&=\frac{\alpha '^s\left(p_2\cdot \epsilon \right){}^s}{3 s!}(-d+4 s+31)
\end{align}
Using \eqref{subsimple}, the complete amplitude for three arbitrary spins with the first leg belonging to the $w=2$ trajectory and two others to the leading ones can be represented as
\begin{align}\label{subsimpleAmp}
\begin{aligned}
    \mathcal{A}^{s_1s_2s_3}&= \int_{\Delta_2} e^{A} \,\Big[ \delta(1-u_2)\delta(u_2-u_1)P_0+\delta(u_2-u_1)P_1 +P_2\Big]\\
    &= \exp[A_0]\, \Big[E_0(A')P_0+E_1(A')P_1+E_2(A')P_2 \Big]
    \end{aligned}
\end{align}
where
\begin{align}
\begin{aligned}
    A &=A_0+ u_1 A' \\
    A_0/\alpha'&=-i p_1\cdot \epsilon _3-i p_3\cdot \epsilon _2-\frac{\epsilon _2\cdot \epsilon _3}{2} \\
    A'/\alpha'&=-2 \left(\epsilon _1\cdot \epsilon _2+\epsilon _1\cdot \epsilon _3+2 i p_2\cdot \epsilon _1\right)
\end{aligned}
\end{align}
and the prefactors $P_i$ are given by
\begin{align}
\begin{aligned}
    P_0/\alpha'&=-2 \left(s_1^2-2 \left(s_2+s_3-3\right) s_1+s_2^2+s_3^2-2 s_3-2 s_2 \left(s_3+1\right)+5\right)\\
    &\qquad +2\alpha '^2\left(p_3\cdot \epsilon _2-p_1\cdot \epsilon _3\right){}^2-\tfrac{8}{3}i\alpha '(s_1+1)p_2\cdot \epsilon _1 \\& \qquad -4\alpha '(s_1+1)(\epsilon _1\cdot \epsilon _2+\epsilon _1\cdot \epsilon _3-\epsilon _2\cdot \epsilon _3)-4i\alpha 'p_3\cdot \epsilon _2(s_1+s_2-s_3+1) \\
    &\qquad -4i\alpha 'p_1\cdot \epsilon _3(s_1-s_2+s_3+1)\\
    P_1/\alpha'&=-\tfrac{8i(d-1)}{3}  p_2\cdot \epsilon _1-16\alpha '[\left(\epsilon _1\cdot \epsilon _2\right){}^2+\left(\epsilon _1\cdot \epsilon _3\right){}^2]-4 (d-1) [\epsilon _1\cdot \epsilon _3+\epsilon _1\cdot \epsilon _2]\\&\qquad +32\alpha '\left(\epsilon _1\cdot \epsilon _3\right)\epsilon _1\cdot \epsilon _2\\
    P_2/{{\alpha'}^2}&=-8 (d-1) \left(\epsilon _1\cdot \epsilon _2-\epsilon _1\cdot \epsilon _3\right){}^2\,,
\end{aligned}
\end{align}
where, after checking that the amplitude is $z$--independent, we have fixed $z_1=0$, $z_2=-1$, $z_3=1$, which choice seems to give the most compact form of the amplitude. To present an example, the amplitude of the lightest member--state of the $w=2$ trajectory, which has spin--$2$, and two gauge bosons reads
\begin{align}
\begin{aligned}
    \mathcal{A}^{s_1=2,s_2=1,s_3=1}&=4 d \epsilon _1\cdot \epsilon _2 \epsilon _1\cdot \epsilon _3 +\frac{8}{3} \alpha'  d \left(\epsilon _1\cdot \epsilon _2 p_1\cdot \epsilon _3 p_2\cdot \epsilon _1+\epsilon _1\cdot \epsilon _3 p_2\cdot \epsilon _1 p_3\cdot \epsilon _2\right)\\
    &\qquad +\frac{2}{3} \alpha'  \left(p_2\cdot \epsilon _1\right){}^2 \left(2 \alpha'  (d+13) p_1\cdot \epsilon _3 p_3\cdot \epsilon _2+(d-39) \epsilon _2\cdot \epsilon _3\right)\,.
    \end{aligned}
\end{align}
It can be easily checked to be gauge invariant for any $d$.

\paragraph{($\boldsymbol{220}$)--amplitudes, etc.} It is also easy to continue replacing leading Regge legs with the $w=2$ trajectory. Indeed, the dressing functions act on separate legs, hence, for example, one only needs to apply \eqref{subsimple} to the second leg as well as the first, to compute $(220)$--amplitudes. The computation is straightforward and we do not give the final answer for it being quite long. Instead, we illustrate it with a couple of particular low--spin amplitudes
\begin{align}
\begin{aligned}
    \mathcal{A}^{s_1=2,s_2=2,s_3=0}&=\frac{1}{9} {\alpha'}^2 \left(d^2+466 d+2249\right) \left(p_2\cdot \epsilon _1\right){}^2 \left(p_3\cdot \epsilon _2\right){}^2\\
    &\quad +\frac{4}{9} \alpha'  \left(17 d^2+194 d+195\right) \epsilon _1\cdot \epsilon _2 p_2\cdot \epsilon _1 p_3\cdot \epsilon _2 \\
    &\quad +\frac{2}{9} \left(15 d^2+162 d+182\right) \left(\epsilon _1\cdot \epsilon _2\right){}^2
\end{aligned}
\end{align}
and
\begin{align}
\begin{aligned}
    \mathcal{A}^{s_1=2,s_2=2,s_3=1}&=\frac{-2}{9}  {\alpha'}^2 \left(d^2+518 d+4953\right) p_1\cdot \epsilon _3 \left(p_2\cdot \epsilon _1\right){}^2 \left(p_3\cdot \epsilon _2\right){}^2\\
    &\quad-\frac{8}{9} \alpha'  \left(17 d^2+246 d+1209\right) \epsilon _1\cdot \epsilon _2 p_1\cdot \epsilon _3 p_2\cdot \epsilon _1 p_3\cdot \epsilon _2\\
    &\quad +\frac{4}{9} \alpha'  \left(d^2+246 d+1209\right) p_2\cdot \epsilon _1 p_3\cdot \epsilon _2 \left(\epsilon _1\cdot \epsilon _3 p_3\cdot \epsilon _2-\epsilon _2\cdot \epsilon _3 p_2\cdot \epsilon _1\right)\\
    &\quad -\frac{4}{3}  \left(5 d^2+54 d+117\right) \epsilon _1\cdot \epsilon _2 \left(\epsilon _1\cdot \epsilon _2 p_1\cdot \epsilon _3+2 \epsilon _1\cdot \epsilon _3 p_3\cdot \epsilon _2+2 \epsilon _2\cdot \epsilon _3 p_2\cdot \epsilon _1\right)\,.
\end{aligned}
\end{align}
The latter, despite involving strange coefficients, can be shown to be gauge invariant for any $d$.

Note that proceeding to $4$--point amplitudes is straightforward using the $N$--point function \eqref{Z_any}.

\paragraph{Closed string amplitudes.} Since closed string correlators on the sphere factorize into (anti)-holomorphic parts, the results obtained in the Section allow us to write down $3$--point amplitudes of closed string trajectories that are obtained by tensoring the treated $w=0$ and/or $w=2$ open string ones, which is the simplest example of the KLT relations \cite{Kawai:1985xq}. For mixed amplitudes of interacting open and closed strings we would have to consider for instance \cite{Garousi:1996ad, Hashimoto:1996kf, Stieberger:2009hq}.

\section{Conclusions and discussion}
\label{sec:conclusions}
Most of the string spectrum is an uncharted territory. In this paper, we have developed a general covariant technique to construct irreducible string states, i.e. those that correspond to elementary particle--like excitations. One advantage of the approach is that states are excavated by \textit{entire} trajectories rather than one by one. There appears also naturally a notion of \textit{complexity} associated with each trajectory: groups of trajectories consist of states with polarization tensors having the symmetries of Young diagrams with a fixed number of rows, with the simplest trajectory corresponding to the first occurrence of the given states in the spectrum, when going up level by level. Recurrent trajectories, hence their member--states, become more and more complicated in the sense of the structure of the dressing functions that act on their simplest versions to create them.

Any particular trajectory, even though it is unbounded in spin, takes advantage of a finite number of modes $\alpha_{-1}, \ldots, \alpha_{-K}$. Irreducible states $|\phi\rangle$ are found by solving the Virasoro constraints, of which the complicated part is $L_{n>0} |\phi\rangle=0$. At a more technical level, our observation is simply that there is a bigger relevant algebra than the one represented on the Fock space by the $L_{n}$. It is $sp(2K)$, which can also be understood as the Howe dual algebra to the Lorentz algebra, with which it commutes. The lowest--weight states of $sp(2K)$ give the simplest solutions to the Virasoro constraints, which already cover infinitely many trajectories, what we call the principal ones. The non--principal trajectories, or recurrent ones, are accessed with the help of trajectory--shifting operators or dressing functions, built out of $sp(2K)$ creation operators. One more advantage of dressing functions is that they allow one to compute generating functions of amplitudes of entire trajectories, rather than of individual states.   

Several extensions and generalizations come to mind. Firstly, it should be possible to extend the technique so as to cover superstring theory, where the Howe dual algebra should be of an $osp(\bullet|\bullet)$--type. Secondly, it is clear that very deep recurrent states are quite complicated and an improvement is needed in order to be able to construct arbitrarily deep  trajectories, even for totally symmetric states. A possible direct application of the techniques developed here is to try to relate the amplitude \cite{Arkani-Hamed:2017jhn} that describes black hole scattering (a different characterization is the unique amplitude with the softest high energy behavior) to string theory, see \cite{Cangemi:2022abk} for such an attempt for the leading Regge trajectory in superstrings.

On the same note of exploring the string spectrum, it is worth mentioning the canonical example of the AdS/CFT correspondence \cite{Maldacena:1997re}. In the tensionless limit, Type IIB string theory on $AdS_5\times S^5$ should be dual to weakly coupled $\mathcal{N}=4$ SYM \cite{Sundborg:2000wp}, hence, the counting of states is equivalent to counting primary operators on the free CFT side \cite{Beisert:2003te, Beisert:2004di}.\footnote{There has also been progress in understanding the tensionless limit of string theory on $AdS_3$ \cite{Gaberdiel:2018rqv,Eberhardt:2019ywk}.} In the free limit, the states are ``1--1'' with cyclic invariant tensor products of the singleton (also called doubleton) representation, see e.g. \cite{Gunaydin:1984fk,Gunaydin:1998sw}. The singleton is a fundamental representation of the higher spin algebra. The latter allows one to repackage the string spectrum into its representations, which consist of infinitely many string states. It is quite amusing that the string spectrum is so simple in such a counter--intuitive limit (in the sense of being ruled by a new, infinite--dimensional, symmetry, the higher spin symmetry), while it appears to be much more complicated already for bosonic strings in flat space. 

\section*{Acknowledgments}
We would like to thank Thomas Basile, Lucile Cangemi, Paolo Di Vecchia, Maxim Grigoriev, Euihun Joung, Axel Kleinschmidt, Renann Lipinski Jusinskas, Pouria Mazloumi, Oliver Schlotterer, Stephan Stieberger, Stefan Theisen and Arkady Tseytlin for very fruitful discussions and correspondence. We  would also like to thank Oliver Schlotterer for very useful comments on an earlier version of the paper. E.S.\ is Research Associate of the Fund for Scientific Research (FNRS), Belgium. The work of C.M. and E.S. was supported by the European Research Council (ERC) under the European Union’s Horizon 2020 research and innovation programme (grant agreement No 101002551) and by the Fonds de la Recherche Scientifique --- FNRS under Grant No. F.4544.21. Both C.M. and E.S. would like to thank the Nordita Institute and program ``Amplifying Gravity at All Scales'' for hospitality while this work was in progress, where also part of our results were presented by C.M.

\appendix

\section{Alpha BRST invariance}
\label{app:alphas}
As is well known, there exists a ``$1$--$1$'' correspondence between vertex operators and physical states, but it is easier to compute amplitudes using the former. Therefore, the Virasoro conditions we studied in the main text can be reformulated for generating functions of $\alpha$--oscillators $\alpha^\mu_n\,$, $\alpha_{-n}^\mu \equiv y_n$. In particular, we can define a generating function $F(\{y_n\})$ of generic states $F \ket{0;p}$. For the latter to be physical, they must be annihilated by the Virasoro generators $L_n$ 
\begin{align} \label{phys_old}
    (L_n -\,\delta_{n,0}) \, F(\{y_m\}) \ket{0;p} &= 0 \quad, \quad \forall \, n \in \mathbb{N} \quad ,\quad m>0\,.
\end{align}
For $n=0$, we obtain from (\ref{phys_old})
\begin{align}\label{mass_lc}
    (N+\alpha' p^2 -1)F&=0\,,
\end{align}
while for $n>0$, we find
\begin{align} \label{general_eq_lc}
   \bigg[\sqrt{2\alpha'} \,n\, p\cdot \frac{\pl}{\pl y_n}+  \frac12\sum_{m=1}^{n-1} m(n-m)\,\frac{\pl^2}{\pl y_{n-m}\cdot \pl y_m}+\sum_{m=0} (n+m+1) y_{m+1}\cdot \frac{\pl}{\pl y_{n+m+1}}\bigg] F&=0\,,
\end{align}
where we have defined
\begin{align}
    \alpha_{m}^\mu \equiv m  \, \eta^{\mu \nu}\frac{\partial}{\partial y_m^\nu} \quad , \quad m >0\,,
\end{align}
since, for example,
\begin{align}
    \alpha_m^\mu \, y_n^\nu  \ket{0;p} = m \, \delta_{mn} \, \eta^{\mu \nu} \ket{0;p} \quad , \quad n,m >0\,.
\end{align}
Moreover, BRST--exact states are created by $L_{-n}$, for example, 
\begin{align} \label{brst_lc}
    L_{-n}&=\tfrac12 \sum_{m=1}^{n-1}y_{m}\cdot y_{n-m}+\sum_{m=0}(m+1)y_{n+m+1}\cdot \frac{\pl}{\pl y_{m+1}}+ \sqrt{2\alpha'}\,p\cdot y_n\,.
\end{align}
The constraints \eqref{mass_lc} and \eqref{general_eq_lc}, as well as the form \eqref{brst_lc}, are equivalent to \eqref{mass_covariant}, \eqref{general_eq_covariant}, upon using the dictionary (\ref{dictionary}), which implies
\begin{align}
    \frac{\partial}{\partial y_m^\mu} = -i \sqrt{2\alpha'} (m-1)! \frac{\partial}{\partial X^{(m)\mu}}
\end{align}
and establishes the equivalence of the definition of the number operator in the oscillator language, \eqref{numberop}, and in the CFT language, \eqref{numbernew}.

\section{Finding trajectories}
\label{app:lightcorn}

\begin{table}
\centering 
\renewcommand{\arraystretch}{1.5}
  \begin{tabular}{ c || c | c | c  }
   $N$ & $gl(d-2)$ tensors & $so(d-2)$ irreps & little group irreps  \\ \hline \hline
   \multirow{2}{*}{$0$}  & $|k\rangle$ & \multirow{2}{*}{$\bullet$} & \multirow{2}{*}{$\bullet$} \\ 
   & $\bullet$ &  \\ \hline
   \multirow{2}{*}{$1$} & $\alpha_{-1}^i|k\rangle$ & \multirow{2}{*}{$\YoungpA$} & \multirow{2}{*}{$\YoungpA$}  \\ 
   & $\YoungpA$  &  \\  \hline
   \multirow{2}{*}{$2$} & $\alpha^{i_1}_{-1}\alpha^{i_2}_{-1}|k\rangle \qquad \alpha^{i}_{-2}|k\rangle$ & \multirow{2}{*}{$\textcolor{red}{\boldpic\YoungpB} \oplus \boldpic\YoungpA \oplus \textcolor{red}{\boldpic\bullet}$} & \multirow{2}{*}{$\boldpic\YoungpB$} \\
   & $\textcolor{red}{\YoungpB} \qquad \qquad \YoungpA$  & & \\ \hline
   \multirow{2}{*}{$3$} & $\alpha^{i_1}_{-1}\alpha^{i_2}_{-1}\alpha^{i_3}_{-1}|k\rangle \quad \alpha^{i_1}_{-2}\alpha^{i_2}_{-1}|k\rangle \quad \alpha^{i}_{-3}|k\rangle$  & \multirow{2}{*}{$\bigg(\textcolor{red}{\boldpic\YoungpC} \oplus \textcolor{blue}{\boldpic\YoungpB} \oplus \textcolor{red}{\boldpic\YoungpA}  \oplus \textcolor{blue}{\bullet} \bigg) \oplus \bigg( \textcolor{blue}{\YoungpAA} \oplus \YoungpA \bigg)$}  & \multirow{2}{*}{$\boldpic\YoungpC \oplus \YoungpAA$} \\
   & $\textcolor{red}{\YoungpC} \qquad \quad  \textcolor{blue}{\YoungpA} \otimes  \textcolor{blue}{\YoungpA}  \qquad  \quad \YoungpA$ & & 
  \end{tabular}
\renewcommand{\arraystretch}{1}
\caption{Open bosonic string, decomposition and recombination per level up to $N=3$.} \label{little_dec}
\end{table}
The unitarity of the bosonic string is easier to prove in the light--cone gauge \cite{Goddard:1972iy, Goddard:1973qh}. After fixing two spacetime coordinates, any state of the spectrum is given by an arbitrary function of the transverse oscillators $\alpha^i_{-n}$, $i=1,...,d-2$, each of which essentially adds $n$ units of energy to a given level $N$. For the lowest levels $N=0$ and $N=1$, precisely $0$ and $1$ oscillator is allowed respectively, so each level contains a single state, the tachyon and a vector respectively, as displayed in table \ref{little_dec}. The vector is consequently the only massless state of the open bosonic string spectrum, with little algebra $so(d-2)$. All other states are massive, so it is rather amusing that a bunch of $gl(d-2)$ tensors recombines into Wigner's little algebra $so(d-1)$ tensors to form any such state. The transverse oscillators produce $gl(d-2)$ tensors since there are no trace constraints; it is then important to apply the level constraint, which will yield the physical states. 

At every $N$, there appears a ``leading'' possible oscillator product
\begin{align} \label{univ_osc}
    &\alpha^{i_1}_{-1}\alpha^{i_2}_{-1}...\alpha^{i_N}_{-1}|k\rangle= \RectARow{4}{$N$}\oplus \RectARow{4}{$N-2$}\oplus ... \oplus\YoungpA/\bullet
\end{align}
where in the RHS we have the trace decomposition of a rank--$N$ $gl(d-2)$ tensor into $so(d-2)$ irreps, as highlighted in red for $N=2,3$ in table \ref{little_dec}. It ends with either a vector or a scalar, depending on whether $N$ is odd or even. It is obvious that \ref{univ_osc} is not capable of delivering a complete set of $so(d-1)$ states. Indeed, the decomposition of a rank--$N$ irreducible $so(d-1)$ tensor into $so(d-2)$ tensors reads
\begin{align}
    \RectARow{4}{$N$}\oplus \RectARow{4}{$N-1$}\oplus ... \oplus\YoungpA \oplus \bullet 
\end{align}
while the irreps with ranks $N-1, N-3, \ldots$ are missing from (\ref{univ_osc}). The spectrum is saved by taking into account oscillators that add more units of energy. In particular, the next possible oscillator product at level $N$ is
\begin{align} \label{osc_save}
    &\alpha^{i_1}_{-1}\alpha^{i_2}_{-1}...\alpha^{i_{N-2}}_{-1}\alpha^{i_{N-1}}_{-2}|k\rangle= \RectARow{4}{$N-2$} \otimes \YoungpA = \RectARow{4}{$N-1$}\oplus \RectBRow{4}{1}{$N-2$}{}
\end{align}
where in the RHS the respective $gl(d-2)$ irreps are displayed, with the $N=3$ example highlighted in blue in table \ref{little_dec}. Now, the $so(d-2)$ decomposition of the first diagram of the RHS of (\ref{osc_save}) supplies the missing states to form the rank--$N$ irrep of $so(d-1)$. This recombination mechanism is illustrated with Young diagrams in bold for $N=2,3$ in table \ref{little_dec}. Next, the second diagram of the RHS of (\ref{osc_save}) requires considering further $\alpha^i_{-3}$, as illustrated in pale font for $N=3$ in table \ref{little_dec}, and so on. For the procedure to work, a neat balance between the energy cost of introducing a new oscillator and the total index structure of the oscillator product is required.

The recombination procedure is algorithmic: one starts with the totally symmetric representation built out of $\alpha_{-1}^i$, decomposes it into $so(d-2)$-irreps, then, one takes the representation with the maximal spin\footnote{Given a Young diagram/weights $(s_1,s_2,...)$, one defines an order by comparing the weights starting from the first one, e.g. $(4,1,...)>(3,2,...)>(3,1,...)$.} upgrades it to $so(d-1)$-irrep with the same Young diagram and subtracts from the set of $so(d-2)$-irreps those obtained by branching this maximal spin $so(d-1)$-irrep, one then adds more oscillators and repeats the cycle. For example, for the states that involve $\alpha_{-1}$ up to $\alpha_{-4}$ one gets
\begin{align}
    \alpha_{-1}\,, \alpha_{-2}&: && \RectARow{4}{$s$}\\
    \alpha_{-1}\,,\alpha_{-2}\,, \alpha_{-3}&: && \RectBRow{4}{1}{$s-2$}{}\\
    \alpha_{-1}\,,\alpha_{-2}\,, \alpha_{-3}\,, \alpha_{-4}&: && \RectARow{4}{$s-2$}\oplus \RectARow{4}{$s-4$}
\end{align}
The decomposition of the string spectrum along these lines is given in Subsection \ref{subsec:zoology}. One simple reason we can start with $(\alpha_{-1})^s$ and add other oscillators step by step is due to the energy cost of $\alpha_{-n}$: the more we introduce the tensors of smaller rank we can build. Therefore, we first make sure that the tensors of the maximal rank are taken care of first and no contradiction can arise when more oscillators are introduced.  The cheap states are easy to point out at any mass level $N$: for each partition $N=\sum_i i\, n_i$ one finds the Young diagram $(n_1,n_2,...)$ among the $so(d-1)$ irreps. Since this diagram does not appear in the decomposition of any other $so(d-1)$ irrep, it has to be present in the physical spectrum. 

It may be entertaining to see to which extent string spectrum is a unique solution. As an assumption, one should ask for a weakly coupled theory that is built on some Fock space. Therefore, in the light-cone gauge one ends up with a set of transverse oscillators. Once the spectrum is mostly massive the recombination problem is present, i.e. the $so(d-2)$ states must result from branching of some $so(d-1)$ irreps at the same mass level. It is not obvious that this can always be done. For example, the leading family built on $\alpha^i_{-1}$ is clearly inconsistent since tensors contracted with $\alpha_{-1}$ cannot be uplifted to $so(d-1)$ ones. One has to add $\alpha_{-2}^i$, which completes the leading Regge trajectory, but creates an incomplete subleading one.    

One spectrum that does not pose any problem at the free level is to have all states massless. The recombination problem is absent and it is also convenient to impose $so(d-2)$ tracelessness to reduce the spectrum further. Obviously, it is possible to keep only the first family of states built on $\alpha_{-1}^i$.  However, this story usually does not end well once interactions are turned on --- there is only a handful of nontrivial theories with massless higher spin fields, see e.g. \cite{Bekaert:2022poo} for a review. 

\section{More on the scattering of general Young diagrams}
\label{app:scYD}

The most general expression that can show up in amplitude calculations corresponds to three (or more for $N$--point amplitudes) polarization tensors of arbitrary Young symmetry $\YYY_k=\YY{s^{k}_1,\ldots,s^{k}_n}$, $k=1,\ldots, N$. The multiplicity of the singlet in the triple tensor product of these $so(d-1)$ representations can be greater than one. It is quite easy to describe what can happen at the $3$--point level. Let the three polarization tensors have the symmetry of $\YYY_k$ diagrams. Indices of each diagram $\YYY_i$ are contracted either with another diagram or with the next momentum $p_{i+1}$ (modulo $3$). The maximal power of the momentum that can be contracted equals the length of the first row. It is convenient to split the Young diagram into rectangular blocks $\{s_j,p_j\}$. Then for every partition $k=\sum_j k_j$ of $k\leq s_1$ such that $k_j\leq s_{j+1}-s_j$ there is a unique contraction of the corresponding momenta. The number of singlets in the tensor product of three diagrams $\YYY_k$ cut in such a way gives the number of independent (atomic) $3$--point amplitudes. The principal states lead to a particular combination of these amplitudes, which is encoded in \eqref{Z_any}. Due to the nature of \eqref{Z_any}, such an amplitude will contain the atomic amplitudes ranging from those with the minimal number of derivatives to the maximal one. What going to the non--principal embedding does is to change the linear combination of these atomic $3$--point amplitudes. The deeper the trajectory the more coefficients get changed.\footnote{An interesting question is whether it is possible to get the atomic amplitudes one by one by choosing the states in a certain way.}
\begin{align*}
\YYY&: &&\parbox{5cm}{\begin{tikzpicture}
    \draw (0,-0.5) -- (5,-0.5) -- (5,-2) -- (3,-2) -- (3,-3.5) -- (1.5, -3.5) -- (1.5, -4.5) -- (0, -4.5) -- (0,-0.5) ;
    \draw [densely dotted] (0,-2) -- (3,-2);
    \draw [densely dotted] (0,-3.5) -- (1.5,-3.5);
    \draw (5,-1.5) -- (3.5, -1.5) --(3.5, -2); 
    \draw (3,-3) -- (2, -3) -- (2, -3.5); 
    \draw (1.5, -4) -- (0.5, -4) -- (0.5, -4.5); 
    \node at (2.5,-0.8) {$s_1$};
    \node at (5.3,-1.3) {$p_1$};
    \node at (4.3,-1.75) {$k_1$};
    \node at (2.5,-3.25) {$k_2$};
    \node at (1,-4.25) {$k_m$};
    \node at (0.8,-4.8) {$s_m$};
    \node at (1.9,-4.0) {$p_m$};
\end{tikzpicture}} && k=\sum k_j\leq s_1\,, \quad k_j\leq s_{j}-s_{j+1}
\end{align*}

\providecommand{\href}[2]{#2}\begingroup\raggedright\endgroup

\end{document}